\documentstyle[12pt]{article}
\def\three_j(#1,#2,#3,#4,#5,#6){\pmatrix{#1 & #2 & #3\cr
					 #4 & #5 & #6\cr}}

\def\qqq{\end{document}}
\def\pmb#1{\setbox0=\hbox{$#1$}%
\kern-.025em\copy0\kern-\wd0
\kern.05em\copy0\kern-\wd0
\kern-.025em\raise.0433em\box0 }

\def\xara(#1,#2,#3,#4){\left(\matrix{#1 & #2\cr #3 & #4\cr}\right)}

\def\six_j(#1,#2,#3,#4,#5,#6){\left\{\matrix{#1 & #2 & #3\cr
					 #4 & #5 & #6\cr}\right\}}
\def\nine_j(#1,#2,#3,#4,#5,#6,#7,#8,#9){\left\{\matrix{#1 & #2 & #3\cr
					#4 & #5 & #6\cr
					 #7 & #8 & #9\cr}\right\}}

\def\Ener(#1,#2){ \sqrt{{#1}^2+{#2}^2} }

\def\overlay#1#2{\setbox0=\hbox{$#1$}\setbox1=\hbox to \wd0{\hss$#2$\hss}#1%
\hskip -1\wd0\copy1}
\def\bold#1{\setbox0=\hbox{$#1$}%
      \kern-.025em\copy0\kern-\wd0
      \kern.05em\copy0\kern-\wd0
      \kern-.025em\raise.0433em\box0 }

\def\S11{S_{11}(1535)}

\def\footnoterule{\kern-3pt \hrule width \hsize \kern2.6pt}

\newcommand{\be}{\begin{equation}}
\newcommand{\ee}{\end{equation}}
\newcommand{\ba}{\begin{eqnarray}}
\newcommand{\ea}{\end{eqnarray}}
\newcommand{\uu}{u({\bf p},s)}
\newcommand{\ubar}{\overline{u}({\bf p'},s')}
\newcommand{\np}{{\bf p}}
\newcommand{\sigvec}{\mbox{\boldmath $\sigma$}}
\newcommand{\etavec}{\mbox{\boldmath $\eta$}}

\newcommand{\kappavec}{\mbox{\boldmath $\kappa$}}
\newcommand{\kvec}{\mbox{\boldmath $\zeta$}}

\newcommand{\gammavec}{\mbox{\boldmath $\gamma$}}

\newcommand{\jvec}{\mbox{\boldmath $J$}}

\newcommand{\bd}[1]{ \mbox{\boldmath $#1$}  }

\newcommand{\nne}{{\bf e}}
\newcommand{\nh}{{\bf h}}
\newcommand{\nK}{{\bf K}}
\newcommand{\nq}{{\bf q}}
\newcommand{\nkappa}{\mbox{\boldmath $\kappa$}}
\newcommand{\nsigma}{\mbox{\boldmath $\sigma$}}

\newcommand{\ntau}{\mbox{\boldmath $\tau$}}
\newcommand{\im}{{\rm Im}\,}
\newcommand{\re}{{\rm Re}\,}

\begin{document}
\begin{center}
{\Large \bf 
%******************************************************************
%
Relativistic Effects in Electromagnetic Meson--Exchange Currents
for One--Particle Emission Reactions
%
%******************************************************************
}\\[1cm]
J.E. Amaro$    ^{1}$, 
M.B. Barbaro$  ^{2}$, 
J.A. Caballero$^{3,4}$, 
T.W. Donnelly$ ^{5}$ and 
A. Molinari$   ^{2}$
\\[0.6cm]
$^{1}${\sl 
Departamento de F\'\i sica Moderna,
Universidad de Granada, 
E-18071 Granada, SPAIN 
}\\[4mm]
$^{2}${\sl 
Dipartimento di Fisica Teorica,
Universit\`a di Torino and
INFN, Sezione di Torino \\
Via P. Giuria 1, 10125 Torino, ITALY 
}\\[4mm]
$^{3}${\sl 
Departamento de F\'\i sica At\'omica, Molecular y Nuclear \\ 
Universidad de Sevilla, Apdo. 1065, E-41080 Sevilla, SPAIN 
}\\[4mm]
$^{4}${\sl 
Instituto de Estructura de la Materia, CSIC \\ 
Serrano 123, E-28006 Madrid, SPAIN 
}\\[4mm]
$^{5}${\sl 
Center for Theoretical Physics, Laboratory for Nuclear Science \\
and Department of Physics\\
Massachusetts Institute of Technology\\ 
Cambridge, MA 02139, USA 
}
\end{center}

\vspace{0.3cm}

%*****************************

\begin{abstract}

%*****************************

Following recent studies of non--relativistic reductions of the
single--nucleon electromagnetic current operator, here we extend the 
treatment to include meson exchange current operators. 
We focus on one--particle emission electronuclear reactions.
In contrast to the 
traditional scheme where approximations are made for the transferred 
momentum, transferred energy and momenta of the initial--state struck 
nucleons, we treat the problem exactly for the transferred energy
and momentum, thus obtaining new current operators which retain
important aspects of relativity not taken into account in the traditional 
non--relativistic reductions. We calculate the matrix elements of our
current operators between the Fermi sphere and a particle--hole
state for several choices of kinematics. We present a comparison between our
results using approximate current operators and those obtained using the 
fully--relativistic operators, as well as with results obtained using the
traditional non--relativistic current operators.

\end{abstract}
%\vspace{0.3in}
%MIT-CTP\#2732  \hfill June 1, 1998

\vfill
\newpage

%*****************************

\section*{1. Introduction}

%*****************************
%REMAINING COMMENTS%%%%%REMAINING COMMENTS%%%%%REMAINING COMMENTS%%%%%

In recent work~\cite{Ama96,Ama96b,Jes98} an improved version of the
single--nucleon electromagnetic current has been studied. There the
so--called ``on--shell form'' of the current was derived as a 
non-relativistic expansion in terms of the dimensionless parameter
$\eta\equiv p/m_N$, where $p$ is the three--momentum of the struck
nucleon (the one in the initial nuclear state to which the virtual
photon in electron scattering reactions is attached) and $m_N$ is
the nucleon mass. Generally (that is, for nucleons in typical
initial--state nuclear wave functions) $p$ lies below a few hundred 
MeV/c and thus $\eta$ is characteristically of order 1/4. Barring
some extreme choice of kinematics such as the selection of extremely
large missing momenta in $(e,e'N)$ reactions --- conditions for which
presently no approach can be guaranteed to work --- an expansion in
powers of $\eta$ is well motivated. Similar arguments do not however
apply equally for other dimensionless scales in the problem. Indeed,
a specific goal of this past work has been to obtain current operators which
are {\em not} expanded in either $\kappa\equiv q/2 m_N$ or $\lambda\equiv
\omega/2 m_N$, where $q$ is the three--momentum and $\omega$ the energy
transferred in the scattering process, since one wishes the formalism to
be applicable at GeV energies where these dimensionless variables are
clearly not small.

Traditionally, many studies have indeed been undertaken assuming
that $\kappa \ll 1$ and $\lambda \ll 1$ aimed of course at treatments
where non--relativistic wave functions are 
employed~\cite{McVoy}--\cite{Adam}. For high--energy conditions the
current operators so obtained are bound to fail, whereas our past work
on the single--nucleon current provides a way to incorporate classes of
relativistic corrections into improved, effective operators for use with
the same non--relativistic wave functions.

Not only the single--nucleon (one--body) current, but also the two--body
meson exchange currents (MEC) have frequently been evaluated using
similar traditional non--relativistic 
expansions \cite{Duba}--\cite{Kohn81} in which $\kappa$ and $\lambda$ are
both treated as being small, together with the assumptions that all
nucleon three--momenta in the problem are small compared with $m_N$. In 
other work \cite{Blun89}--\cite{Dekk92}, relativistic currents have
been used directly in cases where the nuclear modeling permitted.

Our goal in the present work is to extend our previous approach for the
single--nucleon current operators now to include a treatment of 
pion--exchange MEC. We make expansions only in $\eta_i\equiv p_i/m_N$,
where $\{p_i\}$ are the initial--state nucleon three--momenta, whereas
we treat the dependences in the on--shell form exactly for $\kappa$,
$\lambda$ and any high--energy nucleon momenta, specifically for any
nucleons in the final state not restricted to lie within the Fermi sea.
Such new MEC operators may straightforwardly be employed in place of
previous non--relativistic expansions using the same non--relativistic
initial and final nuclear wave functions employed in the past, since
our effective operators incorporate specific classes of relativistic
effects (see the discussions of the single--nucleon current referred to
above).

In the present work, as a first step, we focus on the general form of the
MEC matrix elements for pionic diagrams (the so--called seagull and
pion--in--flight contributions) and plan to extend our treatment to other
diagrams and other meson exchanges in future work. We do not present any
results for electromagnetic response functions, postponing such
discussions until the corresponding correlation effects have been brought
under control (also work in progress). Finally, in the present work we focus on
specific classes of matrix elements, namely those with one
high--energy nucleon in the final state, i.e. one--particle--one--hole (1p--1h)
matrix elements; in subsequent work we shall extend the scope to include 2p--2h
configurations.

The organization of the paper is as follows: After reviewing the treatment of
the single--nucleon current in Sect.~2.1, in Sect.~2.2 we discuss the
new approach for the electromagnetic meson--exchange 
currents that treats the problem exactly for the transferred energy and
transferred momentum. We check the quality of our expansions in powers
of the bound nucleon momenta divided by $m_N$ by calculating the matrix 
elements of the
current operators between the Fermi sphere and a particle--hole state. We
compare with the matrix elements obtained using the full current operators
as well as with the results for the traditional non--relativistic
expansions. These results are presented in Sect.~3 (with reference to
specific details that are covered in an appendix), together 
with a brief discussion of the high--$q$ limits reached by the currents.
Finally in Sect.~4 we summarize our main conclusions. 

%**********************************

\section*{2. Current Operators}

%**********************************

%======================================================
\subsection*{2.1. The Electromagnetic Current Operator}
%======================================================

We start our discussion with the single--nucleon on--shell electromagnetic
current operator and its non--relativistic reduction. Although this 
case has been
already treated in detail in Refs.~\cite{Ama96,Ama96b,Jes98},
here we provide a simpler derivation 
%that is simpler to follow 
%when considering only the lowest--order terms in the non--relativistic 
%expansion (see below); the same approach 
that will be also applied to the case
of the MEC operators for one--particle emission reactions.
The single--nucleon electromagnetic
current reads
\be
J^{\mu}(P's';Ps) = 
\ubar \left[ F_1(Q^2)\gamma^\mu + \frac{i}{2m_N}F_2(Q^2)\sigma^{\mu\nu}Q_\nu 
\right] \uu\ ,
\label{eq1}
\ee
where $P^\mu = (E,\np)$ is the four--momentum of the incident nucleon,
$P'^\mu = (E',\np')$ the four--momentum of the outgoing nucleon and
$Q^\mu = P'^\mu-P^\mu =(\omega,{\bf q})$ 
the transferred four--momentum. The spin projections
for incoming and outgoing nucleons are labeled $s$ and $s'$, respectively.
We follow the conventions of Bjorken and Drell~\cite{BD} for the 
$u$--spinors.
For convenience in the discussions that follow of the scales in the problem 
%and to make the expansion choices transparent 
we introduce the dimensionless variables:
$\etavec = {\bf p}/m_N$, 
$\varepsilon = E/m_N=\sqrt{1+\eta^2}$,
$\lambda = \frac{\omega}{2m_N}$,
$\kappa = \frac{q}{2m_N}$ and
$\tau = -\frac{Q^2}{4m_N^2} = \kappa^2-\lambda^2$.
For the outgoing nucleon, $\etavec'$ and $\varepsilon'$ are defined
correspondingly. 
%Moreover, note
%that the following kinematic relations are satisfied
%\be
%\etavec' =\etavec +2\kappavec
%\label{eq7}
%\ee
%\be
%\varepsilon' =\varepsilon + 2\lambda\ .
%\label{eq8}
%\ee
%Using the notation introduced above, the $u$--spinors 
%are given by~\cite{BD}
%\be 
%\uu = \sqrt{\frac{1+\varepsilon}{2}}
%{1\choose{\frac{1}{1+\varepsilon}\sigvec\cdot\etavec}} \chi_s
%\label{eq9}
%\ee
%normalized to $\overline{u}({\bf p},s) \uu = 1$.

%It is our aim to obtain expressions for the single--nucleon
%electromagnetic current operators $\overline{J}^\mu(P;P')$ that occur
%inside the two--component spin-$\frac{1}{2}$ spinors, {\it viz.}
%\be
%J^\mu(P's';Ps)\equiv \chi^\dagger_{s'}\overline{J}^\mu(P';P)\chi_s\,\, .
%\label{eq10}
%\ee
%The bar over the current distinguishes an operator from its spin matrix
%elements. 
For any general operator whose $\gamma$-matrix form is given by
\be
\Gamma=
\left(
\begin{array}{cc}
\Gamma_{11} & \Gamma_{12}\\
\Gamma_{21} & \Gamma_{22}
\end{array}
\right)
\label{eq11}
\ee
one has $\ubar\Gamma\uu=\chi^\dagger_{s'}\overline{\Gamma}\chi_s$, with the
current operator $\overline{\Gamma}$ given by 
\be
\overline{\Gamma}= \frac{1}{2}\sqrt{(1+\varepsilon)(1+\varepsilon')}
\left(\Gamma_{11}+
\Gamma_{12}\frac{\sigvec\cdot\etavec}{1+\varepsilon}-
\frac{\sigvec\cdot\etavec'}{1+\varepsilon'}\Gamma_{21}-
\frac{\sigvec\cdot\etavec'}{1+\varepsilon'}\Gamma_{22}
\frac{\sigvec\cdot\etavec}{1+\varepsilon}\right)\ .
\label{eq12}
\ee
This general result will be used throughout this work in discussing the
non--relativistic reductions of the various current operators.

An important point in our approach is that we
expand only in powers of the bound nucleon momentum 
$\eta$, not in the transferred momentum $\kappa$ or
the transferred energy $\lambda$. This is a very reasonable approximation
as the momentum of the initial nucleon is relatively low in most
cases, since the typical values of $\eta$ lie below $\eta_F\equiv k_F/m_N$,
where $k_F$ is the Fermi momentum ($\eta_F$ is typically about 1/4). 
However, for those cases corresponding to short--range properties
of the nuclear wave functions it will be necessary to be very careful with
the approximations made. Indeed, for large values of $\eta$ a 
fully--relativistic
approach will likely prove necessary. Expanding up to first order in powers of
$\eta$ we get
$\varepsilon \simeq 1$ and
$\varepsilon' \simeq 1+2\lambda$.
%Other relations that can be obtained easily and will be used later are
%\ba
%\lambda &\simeq& \tau+\kappavec\cdot\etavec \label{eq15}\\
%\kappa^2 &\simeq & \tau(1+\tau+2\kappavec\cdot\etavec) \ . \label{eq16}
%\ea
Thus, the non--relativistic
reductions of the time and space components of the single--nucleon
electromagnetic current operator can be evaluated in a rather simple form.
%\vspace{0.15cm}
%\begin{center}
%\underline{\sl Time Component}
%\end{center}
%\vspace{0.15cm}

Let us consider first the case of the time component. We have
\be
J^0(P's';Ps)=
\ubar J^0 \uu = \chi^\dagger_{s'}\overline{J^0}\chi_s\ ,
\label{eq17}
\ee
with the current operator $J^0=F_1\gamma^0+iF_2\sigma^{0\nu}Q_\nu/2m_N$.
%or, in matrix form,
%\be
%J^0=
%\left(
%\begin{array}{cc}
%F_1  &
%F_2 \sigvec\cdot\kappavec \\
%F_2\sigvec\cdot\kappavec  &  -F_1 
%\end{array}
%\right)\ .
%\label{eq18}
%\ee
Using the
%Introducing this expression in the 
general result given by Eq.~(\ref{eq12}) and
expanding up to first order in $\eta$, it is straightforward
to get the relation
\be
\overline{J^0}\simeq
	\frac{\kappa}{\sqrt{\tau}}G_E
	+\frac{i}{\sqrt{1+\tau}}\left(G_M-\frac{G_E}{2}\right)
	(\kappavec\times\etavec)
	\cdot\sigvec\ ,
\label{eq22}
\ee
where we have introduced 
%\ba
%\overline{J^0} &\simeq &
%	\frac{1}{\sqrt{1+\lambda}}\left\{
%	(1+\lambda)F_1-\left(\tau +\lambda^2\right)F_2+iF_2
%	(\kappavec\times\etavec)\cdot\sigvec \right. \nonumber \\
%& + & \left.
%	\frac{1}{2} (F_1+\lambda F_2) \left[
%	\kappavec\cdot\etavec +i(\kappavec\times\etavec)\cdot\sigvec\right]
%	\right\}\ .
%\label{eq19}
%\ea
%This result can be recast in a simpler form by
%introducing 
the Sachs form factors $G_E=F_1-\tau F_2$ and
$G_M=F_1+F_2$, and have used the relations
% Eq.~(\ref{eq15}). Then, one obtains
%\be
%\overline{J^0} \simeq 
%	\frac{1}{\sqrt{1+\tau+\kappavec\cdot\etavec}}
%	 \left\{
%	\left(1+\tau+\frac{3\kappavec\cdot\etavec}{2}\right)G_E+
%	i\left(G_M-\frac{G_E}{2}\right)(\kappavec\times\etavec)\cdot\sigvec
%	\right\}\ .
%\label{eq20}
%\ee
%The term $[1+\tau+\kappavec\cdot\etavec]^{-1/2}$ can be also expanded 
%in powers of $\eta$. Restricting ourselves to first order and using the
%relation
\ba
\lambda &\simeq& \tau+\kappavec\cdot\etavec \label{eq15}\\
\kappa^2 &\simeq & \tau(1+\tau+2\kappavec\cdot\etavec) \label{eq16}\ .
%\frac{1+\tau+\kappavec\cdot\etavec}{\sqrt{1+\tau}}
%&\simeq &\frac{\kappa}{\sqrt{\tau}}\ ,
%\label{eq21}
\ea
%can finally write for the time component of the electromagnetic current
%\be
%\overline{J^0}\simeq
%	\frac{\kappa}{\sqrt{\tau}}G_E
%	+\frac{i}{\sqrt{1+\tau}}\left(G_M-\frac{G_E}{2}\right)
%	(\kappavec\times\etavec)
%	\cdot\sigvec\ .
%\label{eq22}
%\ee
The expression~(\ref{eq22}) coincides with the leading--order expressions already 
obtained in previous work~\cite{Ama96,Jes98}; in those studies a different 
approach was taken which, while more cumbersome, does yield terms of higher 
order than the ones considered in the present work. 
It is important to remark again that
no expansions have been made in terms of the transferred energy and 
transferred momentum; indeed, $\kappa$, $\lambda$ and $\tau$ may be
arbitrarily large in our approach.

%\vspace{0.15cm}
%\begin{center}
%\underline{\sl Space Components}
%\end{center}
%\vspace{0.15cm}
Let us consider now the case of space components. Thus,
we have ${\bd J}(P's';Ps)=\ubar{\bd J}\uu =
\chi^\dagger_{s'}\overline{\bd J}\chi_s$.
% and the
%matrix form of the vector component for the
%single--nucleon electromagnetic
%current operator is given by
%\be
%{\bd J}=
%\left(
%\begin{array}{cc}
%iF_2 (\sigvec \times \kappavec)  &
%(F_1+\lambda F_2)\sigvec \\
%(-F_1+\lambda F_2)\sigvec  &  iF_2(\sigvec\times \kappavec) 
%\end{array}
%\right)\ .
%\label{eq23}
%\ee
Introducing the matrix form of the vector component for the
single--nucleon electromagnetic current operator in
the general relation 
(\ref{eq12}),
% and following
%a similar procedure to the one applied for the time component, 
one can finally write
\ba
\overline{\bd J} &\simeq & 
	\frac{1}{\sqrt{1+\tau}}\left\{
	iG_M(\sigvec\times\kappavec) 	+
	\left(G_E+\frac{\tau}{2}G_M\right)\etavec + G_E\kappavec \right.
\nonumber \\
& -& 	\left. \frac{G_M}{2(1+\tau)}(\kappavec\cdot\etavec)\kappavec-
\frac{iG_E}{2(1+\tau)}(\sigvec\times\kappavec)\kappavec\cdot\etavec
\right. \nonumber \\
	&-& \left. i\tau(G_M-G_E/2)(\sigvec\times\etavec)+
	\frac{i(G_M-G_E)}{2(1+\tau)}(\kappavec\times\etavec)
	\sigvec\cdot\kappavec \right\}\ ,
\label{eq24}
\ea
where we have used the relations given by 
Eqs.~(\ref{eq15},\ref{eq16}).
% and have expanded
%the term $[1+\tau+\kappavec\cdot\etavec]^{-1/2}$ in powers of $\eta$ (up
%to first order). The general vector identity
%\be
%(\sigvec\cdot\etavec')(\sigvec\times\kappavec)(\sigvec\cdot\etavec) =
%-(\etavec\cdot\etavec')(\sigvec\times\kappavec)
%- (\kappavec\times\etavec)\sigvec\cdot(\etavec+\etavec')
%+2i\kappavec\times(\kappavec\times\etavec) 
%\label{eq25}
%\ee
%has also been used. Note that this identity reduces to
%\be
%(\sigvec\cdot\etavec')(\sigvec\times\kappavec)(\sigvec\cdot\etavec) \simeq
%-2\left[(\etavec\cdot\kappavec)(\sigvec\times\kappavec)
%+ (\kappavec\times\etavec)(\sigvec\cdot\kappavec)
%-i\kappavec\times(\kappavec\times\etavec)\right]
%\label{eq26}
%\ee
%in the non--relativistic expansion up to linear order in $\eta$.

In order to compare with the previous work~\cite{Jes98},
we write the expression for the transverse component of the current, {\em i.e.},
$\overline{\bd J}^\perp =\overline{\bd J}-
\frac{\overline{\bd J}\cdot\kappavec}{\kappa^2}
\kappavec$. After some algebra 
%and making use of the
%vector identity
%\be
%\kappavec\left[(\sigvec\times\etavec)\cdot\kappavec \right] =
%-(\kappavec\cdot\sigvec)(\kappavec\times\etavec) 
%+ (\kappavec\cdot\etavec)(\kappavec\times\sigvec)
%+\kappa^2(\sigvec\times\etavec)\ ,
%\label{eq27}
%\ee
we get the final result
\ba
\overline{\bd J}^\perp &\simeq&
\frac{1}{\sqrt{1+\tau}}\left\{
	iG_M(\sigvec\times\kappavec)+
	\left(G_E+\frac{\tau}{2}G_M\right)
	\left(\etavec
	-\frac{\kappavec\cdot\etavec}{\kappa^2}\kappavec\right) 
\right. \nonumber \\
&-& \left.
	\frac{iG_M}{1+\tau}
	(\sigvec\times\kappavec)\kappavec\cdot\etavec 
+	\frac{iG_M}{2(1+\tau)}(\etavec\times\kappavec)
	\sigvec\cdot\kappavec \right\}\ .
\label{eq28}
\ea
It is straightforward to prove that this expression coincides with the
result given by Eq.~(25) in Ref.~\cite{Jes98} for an expansion in powers
of $\eta$ up to first order.
%. One should only
%take into account the two following relations that are valid for an
%expansion in powers of $\eta$ up to first order
%\ba
%\frac{\sqrt{\tau}}{\kappa}&\simeq & \frac{1}{\sqrt{1+\tau}}
%\left(1-\frac{\kappavec\cdot\etavec}{1+\tau}\right)
%\nonumber  \\
%\frac{\lambda}{2\kappa^2}&\simeq & \frac{1}{2(1+\tau)}
%\left[1+\frac{1-\tau}{\tau(1+\tau)}\kappavec\cdot\etavec\right]\ .
%\label{eq29}
%\ea

Therefore, as can be seen from Eqs.~(\ref{eq22},\ref{eq28}), at linear order in
$\eta$ we retain the 
spin--orbit part of the charge and one of the relativistic corrections to
the transverse current, the first--order convective spin--orbit term.
It is also important to remark here that the 
current operators given by Eqs.~(\ref{eq22},\ref{eq24}) satisfy the property of
current conservation $\lambda J_0 = \kappavec\cdot{\bd J}$.
Finally, it is also interesting to quote the results obtained in the 
traditional non--relativistic
reduction~\cite{Jes98},\cite{BGP80}--\cite{Adam}, where it is assumed that
$\kappa <<1$ and $\lambda << 1$: 
\ba
\overline{J^0}_{nonrel} &=& G_E \nonumber \\
\overline{\bd J}_{nonrel}^\perp &=& -iG_M[\kappavec \times \sigvec]+
	G_E\left[\etavec-\left(\frac{\kappavec\cdot\etavec}{\kappa^2}
	\right)\kappavec\right]\ .
\label{eq30}
\ea
Note that this traditional
non--relativistic reduction contains both terms of zeroth
and first order in $\eta$, {\em i.e.}, the convection
current, and is therefore not actually of 
lowest order in $\eta$. 
%Moreover, note that
%$\kappavec\cdot\overline{\bd J}_{nonrel}=0$, {\it i.e.,} this
%non--relativistic current is not conserved.

The present expansion for the electromagnetic current operator of the nucleon
was first checked in Ref.~\cite{Ama96}, where the inclusive
longitudinal and transverse responses of a non--relativistic 
Fermi gas were found to agree with the exact relativistic
result within a few percent if one uses 
relativistic kinematics
when computing the energy of the ejected nucleon.
Recently the same  expansion has been tested 
with great success by comparing with the relativistic
exclusive polarized responses for the $^2$H$(e,e'p)$ reaction 
at high momentum transfers~\cite{Jes98}. 
This relativized current has also been applied to the calculation of 
inclusive and exclusive responses that arise in the 
scattering of polarized electrons  
from polarized nuclei~\cite{Ama96b,Ama97}. 

We see that the expansion of the current {\em to first order} in the variable 
$\eta=p/m_N$ yields quite simple expressions;
moreover the various surviving pieces of the relativized current
(i.e., charge and spin--orbit in the 
longitudinal and magnetization and convection in the transverse) differ
from the traditional non--relativistic expressions
only by multiplicative $(q,\omega)$-dependent factors
such as $\kappa/\sqrt{\tau}$ or $1/\sqrt{1+\tau}$, and therefore
are easy to implement in already existing non--relativistic 
models. In the next section we  perform a similar
expansion for the MEC and later return to check our results through 
direct comparisons with the exact relativistic matrix elements.

%======================================================
\subsection*{2.2. Meson--Exchange Currents}
%======================================================

Once the procedures for expanding the single--nucleon 
electromagnetic current are fixed, it is clear how to proceed to obtain 
relativistic expansions for the meson--exchange currents. In the present
work we begin by focusing on pion exchange MEC effects, leaving the 
treatment of other mesons to future work. 
Following the ideas and methods developed in the 
previous section, our main aim here is to get new non--relativistic reductions 
for MEC treating the problem of the transferred energy and
transferred momentum as above, namely in an un-expanded form while expanding
only in the initial nucleon momenta. In this way the expressions 
obtained will retain important
aspects of relativity not included in the traditional non--relativistic
MEC used throughout the literature. 

Let us consider the meson exchange current 
operator $J^{MEC}_\mu$. For definiteness we focus on one--particle
emission reactions where the matrix element of
$J^{MEC}_\mu$ taken between the Fermi sphere and a particle--hole state,
namely 
\be
\langle ph^{-1}|J^{MEC}_\mu|F\rangle =
	\sum_{s',t'}\sum_{h'<k_F}\left[
	\langle ph'|J^{MEC}_\mu|hh'\rangle -
	\langle ph'|J^{MEC}_\mu|h'h\rangle \right]\ ,
\label{eq31}
\ee
$k_F$ being the Fermi momentum, is the relevant one.
Since, as seen below when the pion--exchange
operators are given, the currents being considered all have isospin
dependences of the form $\tau_a(1)\tau_b(2)$, the first term (direct term) 
vanishes, 
\be
\sum_{t'}\langle t_pt'|\tau_a(1)\tau_b(2)|t_ht'\rangle =
	\langle t_p|\tau_a|t_h\rangle \sum_{t'}
	\langle t'|\tau_b|t'\rangle  =0\ .
\label{eq32}
\ee
Therefore, the only remaining term is the exchange term, and we can simply
write
\be
\langle ph^{-1}|J^{MEC}_\mu |F\rangle = -\sum_{s',t'}
\sum_{h'<k_F}\langle ph'|J^{MEC}_\mu|h'h\rangle\ .
\label{eq33}
\ee
In what follows we will be interested in the evaluation of the
particle--hole matrix elements $\langle ph'|J^{MEC}_\mu|h'h\rangle$ 
and their {\sl new} non--relativistic expressions. As in our treatment
of the single--nucleon current, it is convenient to express the results
in terms of spin matrix elements of particular operators 
%(acting in spin space; compare Eq.~(\ref{eq10}))
:
\be
\langle ph'|J^{S,P}_\mu|h'h\rangle
  \equiv \chi^\dagger_{s'_1}\chi^\dagger_{s'_2} \overline{J}_{\mu}^{S,P}(1,2)
  \chi_{s_1}\chi_{s_2}  \ ,
\label{eq33a}
\ee
where $S$($P$) denotes the seagull (pion--in--flight) contributions
to the MEC, as shown in Fig.~1.

%======================================================
\subsubsection*{2.2.1. Seagull Current Operator}
%======================================================

The relativistic seagull current operator is given by
\ba
J_{\mu}^{S}(Q) &=& \frac{f^2}{V^2 m_\pi^2}i\epsilon_{zab}
             \bar{u}({\bf p'}_1,s'_1)\tau_a\gamma_5\not{\!\!K_1}u({\bf p}_1,s_1)
             \frac{F_1^V}{K_1^2-m_\pi^2}
             \bar{u}({\bf p'}_2,s'_2)\tau_b\gamma_5\gamma_{\mu}u({\bf p}_2,s_2)
\nonumber
\\
             &+& (1 \leftrightarrow 2)\ ,
\label{e34}
\ea
where $V$ is the volume enclosing the system and the different nucleon
kinematic variables are: 
$P_1=(E_1,{\bf p}_1)$, $P'_1=(E'_1,{\bf p}'_1)$, 
$P_2=(E_2,{\bf p}_2)$ and $P'_2=(E_2,{\bf p}'_2)$ (see Fig.~1). 
The four--momentum of the exchanged pion is 
$K_1=(E_{k_1},{\bf k}_1)$ (with $K_2$ likewise) and its mass is $m_\pi$. The 
terms $f$ and $F_1^V$ represent the pion--nucleon coupling and
pseudovector form factor, respectively. 
Note that the following kinematic relations are satisfied for the two diagrams
involved 
\ba
\etavec'_1 &=& \etavec_2 + 2\kappavec \label{eq35}\\
\etavec'_2 &=& \etavec_1 \label{eq36}\\
\kvec_1 &=& \etavec'_1-\etavec_1
=\etavec_2-\etavec_1+2\kappavec
\label{eq37a}
\\
\kvec_2 &=& \etavec'_2-\etavec_2
=\etavec_1-\etavec_2\ ,
\label{eq37b}
\ea
where we follow the general notation introduced in
%Eqs.~(\ref{eq2}--\ref{eq6}) 
Section 2.1 and have also defined
$\kvec_{1,2}\equiv {\bf k}_{1,2}/m_N$ with ${\bf k}_{1,2}$ the three--momenta
of the exchanged pions.

The particle--hole matrix element of the seagull current is then
given by
\ba
\langle ph'|J_\mu^S|h'h\rangle &=&
\frac{f^2}{V^2 m_\pi^2}i\epsilon_{zab}\langle t_p|\tau_a|t_{h'}\rangle
		\langle t_{h'}|\tau_b|t_h\rangle \nonumber \\
&\times &\left\{    \bar{u}({\bf p'}_1,s'_1)\gamma_5\not{\!\!K_1}u({\bf p}_1,
s_1) 
             \frac{F_1^V}{K_1^2-m_\pi^2}
             \bar{u}({\bf p'}_2,s'_2)\gamma_5\gamma_{\mu}u({\bf p}_2,s_2)
\right. \nonumber \\
&-& \left.
    \bar{u}({\bf p'}_2,s'_2)\gamma_5\not{\!\!K_2}u({\bf p}_2,s_2)
             \frac{F_1^V}{K_2^2-m_\pi^2}
             \bar{u}({\bf p'}_1,s'_1)\gamma_5\gamma_{\mu}u({\bf p}_1,s_1)
\right\}\ ,
\label{eq38}
\ea
where now $\np'_1=\np$ is the momentum of the ejected particle
above the Fermi sea, which can be large for large 
momentum transfer, while $\np_2=\nh$ is the
momentum of the bound nucleon before the interaction
(related to the missing momentum in $(e,e'p)$ reactions), 
which can be considered small compared to the nucleon mass.
Finally, 
$\np'_2=\np_1=\nh'$ is the intermediate momentum of the bound nucleon 
interacting with the ejected nucleon  by pion exchange. 
Therefore $\np'_2$ and $\np_1$ are small compared with the nucleon 
mass. In dimensionless terms, we can safely expand in $\etavec_1$, 
$\etavec_2$ and hence $\etavec'_2$, whereas we cannot in general expand
in $\etavec'_1$. Moreover, $\kvec_2$ is small, whereas $\kvec_1$ can be large. 
By analogy with the single--nucleon electromagnetic current, let us
now proceed to the evaluation of the time and space components separately.
\vspace{0.15cm}
\begin{center}
\underline{\sl Time Component}
\end{center}
\vspace{0.15cm}
Using the general result given by Eq.~(\ref{eq12}) and the matrices 
$\gamma_5\not{\!\!K}_1$ and $\gamma_5\gamma_0$ we can write
\ba
& & \langle ph'|J_0^S|h'h\rangle =
-\frac{f^2}{4V^2m_\pi^2}i\epsilon_{zab}\langle t_p|\tau_a|t_{h'}\rangle
		\langle t_{h'}|\tau_b|t_h\rangle
\sqrt{(1+\varepsilon_1)(1+\varepsilon_2)(1+\varepsilon'_1)
(1+\varepsilon'_2)} \nonumber \\
&\times& \left\{\chi^\dagger_{s'_1}\left[m_N\sigvec\cdot\kvec_1-
E_{k_1}\frac{\sigvec\cdot\etavec_1}{1+\varepsilon_1}-
E_{k_1}\frac{\sigvec\cdot\etavec'_1}{1+\varepsilon_1'}+
m_N\frac{\sigvec\cdot\etavec_1'}{1+\varepsilon'_1}
(\sigvec\cdot\kvec_1)
\frac{\sigvec\cdot\etavec_1}{1+\varepsilon_1}\right]\chi_{s_1}
\right. \nonumber \\
&\times& \left. \frac{F_1^V}{K_1^2-m_\pi^2}\chi^\dagger_{s'_2}\left[
\frac{\sigvec\cdot\etavec_2}{1+\varepsilon_2}+
\frac{\sigvec\cdot\etavec_2'}{1+\varepsilon'_2}\right]\chi_{s_2}
\right. \nonumber \\
&-& \left. 
\chi^\dagger_{s'_1}\left[
\frac{\sigvec\cdot\etavec_1}{1+\varepsilon_1}+
\frac{\sigvec\cdot\etavec_1'}{1+\varepsilon'_1}\right]\chi_{s_1}
\frac{F_1^V}{K_2^2-m_\pi^2} \right. \nonumber \\
&\times& \left. 
\chi^\dagger_{s'_2}\left[m_N\sigvec\cdot\kvec_2-
E_{k_2}\frac{\sigvec\cdot\etavec_2}{1+\varepsilon_2}-
E_{k_2}\frac{\sigvec\cdot\etavec_2'}{1+\varepsilon_2'}+
m_N\frac{\sigvec\cdot\etavec_2'}{1+\varepsilon'_2}
(\sigvec\cdot\kvec_2)
\frac{\sigvec\cdot\etavec_2}{1+\varepsilon_2}\right]\chi_{s_2}
\right\}\ ,
\label{eq39}
\ea
where $E_{k_1}(E_{k_2})$ is, as mentioned previously, the energy of the pion. 

We note that the resulting expansion for the  MEC
should be used together with the single--nucleon current,
which has been developed to first order in $\eta$. 
Therefore, in order to be consistent, we want to perform the
expansion of the MEC also to first order in the corresponding
small quantities $\{\etavec_1, \etavec_2, \etavec'_2, \kvec_2\}$, whereas
$\{\etavec'_1, \kappavec, \kvec_1\}$ are treated exactly.
Using the kinematic relations given
in Eqs.~(\ref{eq35}--\ref{eq37b}), the following
non--relativistic kinematic reductions are involved (up to first order
in the small quantities)
\ba
\varepsilon_1&\simeq& \varepsilon_2 \simeq \varepsilon'_2 \simeq 1 
\label{eq40}\\
\varepsilon'_1&\simeq & 1+2\lambda \label{eq41}\\
E_{k_1} &\simeq &  2m_N\lambda \label{eq42}\\
E_{k_2} &\simeq &  0\ ,
\label{eq43}
\ea
where we follow the notation introduced in 
%Eqs.~(\ref{eq2}--\ref{eq6})
Section 2.1. Using these non--relativistic expansions
and the kinematic relations given by Eqs.~(\ref{eq35}--\ref{eq37b}),
it is straightforward to obtain 
\ba
\overline{J}_{0}^{S}(1,2) &=&
\frac{{\cal F}}{2 \sqrt{1+\lambda}} \left\{
\frac{\sigvec_1\cdot\left(\kvec_1-\lambda\etavec_1\right)
\sigvec_2\cdot(\etavec_2+\etavec'_2)}{K_1^2-m_\pi^2} 
\right. \nonumber \\
&-& \left. 
\frac{\sigvec_1\cdot\left[(1+\lambda)\etavec_1+\etavec'_1\right]
\sigvec_2\cdot\kvec_2}{K_2^2-m_\pi^2} 
\right\}\ ,
\label{eq44}
\ea
having defined
\be
{\cal F} = -\frac{f^2m_N}{V^2m_\pi^2}
i\epsilon_{zab}\langle t_p|\tau_a|t_{h'}\rangle 
\langle t_{h'}|\tau_b|t_h \rangle F_1^V\ .  
\label{eq44a}
\ee
We can still simplify this expression by using the relation
$\lambda \simeq \tau+\kappavec\cdot\etavec_2$ (valid up to first order in
powers of $\eta_2$): the result is
\ba
\overline{J}_{0}^{S}(1,2) &=&
\frac{{\cal F}}{2 \sqrt{1+\tau}} \left\{
\frac{\sigvec_1\cdot\left(\kvec_1-\tau\etavec_1\right)
\sigvec_2\cdot(\etavec_2+\etavec'_2)}{K_1^2-m_\pi^2} 
\right. \nonumber \\
&-& \left. 
\frac{\sigvec_1\cdot\left[(1+\tau)\etavec_1+\etavec'_1\right]
\sigvec_2\cdot\kvec_2}{K_2^2-m_\pi^2} 
\right\}\ .
\label{eq45}
\ea

Examining this result, we see that we have retained some terms that should 
actually be neglected because they
give 2$^{nd}$--order contributions: this is the case for the factor
$\tau(\sigvec_1\cdot\etavec_1)$
which is multiplied by $\sigvec_2\cdot(\etavec_2+\etavec'_2)$ and
$\sigvec_2\cdot\kvec_2$. When these contributions to the current are omitted,
then the comparison between our result and the traditional non--relativistic
expression~\cite{Alb90},~\cite{Alb93} becomes more straightforward. 
Moreover note that if one neglects the term
$\tau(\sigvec_1\cdot\etavec_1)$, then
one recovers for the time component of the seagull current an expression that
is similar to the traditional
non--relativistic reduction~\cite{Alb90} {\em except} for the common factor 
$1/\sqrt{1+\tau}$, which accordingly incorporates 
important aspects of relativity
not considered in the traditional non--relativistic reduction.
This result is similar
to the discussion given in Ref.~\cite{Ama96} for the case of the
single--nucleon electromagnetic current. 
Finally, note that, strictly speaking, in order to
be consistent with the non--relativistic expansion, the
contribution of $\sigvec_1\cdot\etavec_1$ should be also neglected. 
Therefore, our final expression
for the non--relativistic reduction of the seagull current can be
written in the form
\be
\overline{J}_{0}^{S}(1,2) =
\frac{{\cal F}}{2 \sqrt{1+\tau}} 
\left\{
\frac{\left(\sigvec_1\cdot\kvec_1\right)\sigvec_2\cdot(\etavec_2+\etavec'_2)}
{K_1^2-m_\pi^2}
-\frac{\left(\sigvec_1\cdot\etavec'_1\right)\left(\sigvec_2\cdot\kvec_2\right)}
{K_2^2-m_\pi^2}
\right\}\ .
\label{eq45a}
\ee 

In order to obtain a truly first--order expansion of the current it is 
convenient to re-express the momenta involved in Eq.~(\ref{eq45}) in terms of
the momentum transfer $\kappavec$, which can in principle be large, and the 
nucleon momenta $\etavec_1$, $\etavec_2$, which lie below the Fermi surface and
are kept as the parameters of the expansion.
By inserting Eqs.~(\ref{eq35}--\ref{eq37b}) into Eq.~(\ref{eq45})
and keeping only terms linear in $\etavec_1$, $\etavec_2$, one gets (note that
we keep the full pion propagators) 
\be
\overline{J}_{0}^{S}(1,2) = {\cal F} 
\frac{\sigvec_1\cdot\kappavec}{\sqrt{1+\tau}} 
\left[
\frac{\sigvec_2\cdot(\etavec_1+\etavec_2)}{K_1^2-m_\pi^2} 
-
\frac{\sigvec_2\cdot(\etavec_1-\etavec_2)}{K_2^2-m_\pi^2} 
\right]\ .
\label{bai1}
\ee

Finally, it is also interesting to examine the 
limit $\eta_F\rightarrow 0$, since this will provide some understanding of
how the MEC effects are expected to evolve in going from light 
($\eta_F$ very small) to heavy nuclei ($\eta_F\cong 0.29$). 
Obviously, in this case the following
relations are satisfied
\ba
& & \etavec_1 = \etavec_2 = \etavec'_2 =\kvec_2=0 \label{eq47}\\
& & \kvec_1 = \etavec'_1= 2\kappavec
\label{eq48}
\ea
and therefore the seagull current simply reduces to
\be
\left[\overline{J}_{0}^{S}(1,2) \right]_{\eta_F\rightarrow 0}=0\ .
\label{eq49}
\ee
This is a consequence of the fact
that the time component of the seagull current is
of first order in the small variables involved
or, in other words, it is of $O(\eta_F)$.

\vspace{0.15cm}
\begin{center}
\underline{\sl Space Components}
\end{center}
\vspace{0.15cm}

The particle--hole matrix element is given by
\ba
\langle ph'|\jvec^S|h'h\rangle &=&
\frac{f^2}{V^2m_\pi^2}i\epsilon_{zab}\langle t_p|\tau_a|t_{h'}\rangle
	\langle t_{h'}|\tau_b|t_h\rangle \nonumber \\
&\times & \left\{    \bar{u}({\bf p'}_1,s'_1)\gamma_5\not{\!\!K_1}u({\bf p}_1,
s_1) 
             \frac{F_1^V}{K_1^2-m_\pi^2}
             \bar{u}({\bf p'}_2,s'_2)\gamma_5\gammavec u({\bf p}_2,s_2)
\right. \nonumber \\
&-& \left.
    \bar{u}({\bf p'}_2,s'_2)\gamma_5\not{\!\!K_2}u({\bf p}_2,s_2)
             \frac{F_1^V}{K_2^2-m_\pi^2}
             \bar{u}({\bf p'}_1,s'_1)\gamma_5\gammavec u({\bf p}_1,s_1)
\right\}\ .
\label{eq50}
\ea
Using again the general relation (12) for the matrix
forms for $\gamma_5\gammavec$ and $\gamma_5\not{\!\!K_{1,2}}$, one can write
\ba
& & \langle ph'|\jvec^S|h'h\rangle \nonumber \\
&=&-\frac{f^2}{4V^2m_\pi^2}i\epsilon_{zab}\langle t_p|\tau_a|t_{h'}\rangle
	\langle t_{h'}|\tau_b|t_h\rangle 
\sqrt{(1+\varepsilon_1)(1+\varepsilon_2)(1+\varepsilon'_1)
(1+\varepsilon'_2)} \nonumber \\
&\times& \left\{\chi^\dagger_{s'_1}\left[m_N\sigvec\cdot\kvec_1-
E_{k_1}\left(\frac{\sigvec\cdot\etavec_1}{1+\varepsilon_1}+
\frac{\sigvec\cdot\etavec'_1}{1+\varepsilon_1'}\right)+
m_N\frac{\sigvec\cdot\etavec_1'}{1+\varepsilon'_1}
\sigvec\cdot\kvec_1
\frac{\sigvec\cdot\etavec_1}{1+\varepsilon_1}\right]\chi_{s_1}
\right. \nonumber \\
&\times& \left. \frac{F_1^V}{K_1^2-m_\pi^2}\chi^\dagger_{s'_2}\left[
\sigvec+\frac{\sigvec\cdot\etavec'_2}{1+\varepsilon'_2}\sigvec
\frac{\sigvec\cdot\etavec_2}{1+\varepsilon_2}\right]\chi_{s_2}
\right. \nonumber \\
&-& \left. 
\chi^\dagger_{s'_1}\left[\sigvec+
\frac{\sigvec\cdot\etavec'_1}{1+\varepsilon'_1}\sigvec
\frac{\sigvec\cdot\etavec_1}{1+\varepsilon_1}\right]\chi_{s_1}
\frac{F_1^V}{K_2^2-m_\pi^2} \right. \nonumber \\
&\times& \left. 
\chi^\dagger_{s'_2}\left[m_N\sigvec\cdot\kvec_2-
E_{k_2}\left(\frac{\sigvec\cdot\etavec_2}{1+\varepsilon_2}+
\frac{\sigvec\cdot\etavec_2'}{1+\varepsilon_2'}\right)+
m_N\frac{\sigvec\cdot\etavec_2'}{1+\varepsilon'_2}
\sigvec\cdot\kvec_2
\frac{\sigvec\cdot\etavec_2}{1+\varepsilon_2}\right]\chi_{s_2}
\right\}\ .
\label{eq51}
\ea

We can now make use of the non--relativistic reductions as explained in
the previous section in connection with the time component. Then,
considering only terms up to first order in the small quantities in which we
are expanding, after some algebra one finally gets 
\be
\overline{\jvec}^{S}(1,2) \simeq {\cal F}
\left\{ \frac{1}{\sqrt{1+\lambda}}
\frac{\sigvec_1\cdot\left(\kvec_1-\lambda\etavec_1\right)\sigvec_2} 
{K_1^2-m_\pi^2} 
- \sqrt{1+\lambda}
\frac{\sigvec_1\left(\sigvec_2\cdot\kvec_2\right)}
{K_2^2-m_\pi^2} 
\right\}\ .
\label{eq52}
\ee
This expression can be further simplified by using the relation
$\lambda \simeq  \tau +\kappavec\cdot\etavec_2$ and expanding
the terms $1/\sqrt{1+\lambda}$ and
$\sqrt{1+\lambda}$ in powers of $\eta_2$ (up to first order). 
The final result is
\be
\overline{\jvec}^{S}(1,2) \simeq {\cal F}
\left\{ \frac{1}{\sqrt{1+\tau}}
\frac{\sigvec_1\cdot\left[
\left(1-\frac{\kappavec\cdot\etavec_2}{2(1+\tau)}\right)\kvec_1
-\tau\etavec_1\right]\sigvec_2} 
{K_1^2-m_\pi^2} 
- \sqrt{1+\tau}
\frac{\sigvec_1\left(\sigvec_2\cdot\kvec_2\right)}
{K_2^2-m_\pi^2} 
\right\}\ .
\label{eq53}
\ee
It is important to note that
neglecting the term $(\kappavec\cdot\etavec_2)/[2(1+\tau)]$ compared to 1, and
$\tau(\sigvec_1\cdot\etavec_1)$ compared to
$\sigvec_1\cdot\kvec_1$ (good approximations --- see the next section)
one simply recovers
the traditional non--relativistic expression~\cite{Alb90} except for the factors
$1/\sqrt{1+\tau}$ and $\sqrt{1+\tau}$ that multiply the
contributions given by the
two diagrams involved. As in the case of the time component,
this result indicates that important relativistic effects can be simply 
accounted for by these multiplicative terms. In next section we shall present
results for a wide choice 
of kinematics showing the validity of the obtained expressions. 
By inserting Eqs.~(\ref{eq35}--\ref{eq37b}) into Eq.~(\ref{eq53}) and expanding
up to first order in $\etavec_1$, $\etavec_2$, one gets
\ba
\overline{\jvec}^{S}(1,2) &=& {\cal F} 
\left\{ 
\left[ 
\frac{2\sigvec_1\cdot\kappavec}{\sqrt{1+\tau}}
\left(1-\frac{\kappavec\cdot\etavec_2}{2(1+\tau)}\right) +
\frac{\sigvec_1\cdot\etavec_2}{\sqrt{1+\tau}} -
\sqrt{1+\tau} \sigvec_1\cdot\etavec_1
\right]
\frac{\sigvec_2}{K_1^2-m_\pi^2} 
\right.
\nonumber
\\
&-&
\left.
\sqrt{1+\tau}
\frac{\sigvec_1}{K_2^2-m_\pi^2} 
\sigvec_2\cdot(\etavec_1-\etavec_2)
\right\}\ .
\label{bai2}
\ea

Finally, in the limit $\eta_F\rightarrow 0$, one obtains
\be
\left[\overline{\jvec}^{S}(1,2) \right]_{\eta_F\rightarrow 0} 
\simeq \frac{2{\cal F}}{\sqrt{1+\tau}}
\frac{(\sigvec_1\cdot\kappavec)\sigvec_2}{Q^2-m_\pi^2}\ ,
\label{eq54}
\ee
which shows that the space components of the seagull current are of $O(1)$ and 
contribute even for nucleons at rest, as do the charge and magnetization
pieces of the one--body current.

%======================================================
\subsubsection*{2.2.2. Pion--in--flight Current Operator}
%======================================================

The relativistic pion--in--flight current operator reads
\be
J_{\mu}^{P}(Q)= \frac{f^2 F_\pi}{V^2m_\pi^2}i\epsilon_{zab}
             \frac{(K_1-K_2)_\mu}{(K_1^2-m_\pi^2)(K_2^2-m_\pi^2)}
             \overline{u}(\np'_1)\tau_a\gamma_5\not{\!\!K_1}u(\np_1)
             \overline{u}(\np'_2)\tau_b\gamma_5\not{\!\!K_2}u(\np_2)\ ,
\label{eq55}
\ee
where the kinematic variables are defined in Fig.~1 and where the kinematic
relationships given in Eqs.~(\ref{eq35}--\ref{eq37b}) for the seagull 
diagram are again satisfied. 
In order to preserve gauge invariance, we choose the electromagnetic pion form
factor to be $F_\pi=F_1^V$ (see Ref.~\cite{Mathiot}).

The exchange particle--hole matrix element is given by
\ba
& & \langle p h'|J_\mu^P|h'h\rangle = 
    \frac{f^2F_1^V}{V^2m_\pi^2}i\epsilon_{zab}\langle t_p|\tau_a|t_{h'}\rangle 
	\langle t_{h'}|\tau_b|t_h\rangle 
	\frac{(K_1-K_2)_\mu}{(K_1^2-m_\pi^2)(K_2^2-m_\pi^2)} \nonumber \\
&\times&	\overline{u}(\np'_1)\gamma_5\not{\!\!K_1}u(\np_1)
             \overline{u}(\np'_2)\gamma_5\not{\!\!K_2}u(\np_2)\ .
\label{eq59}
\ea
Again using Eq.~(\ref{eq12}) and the matrix forms of $\gamma_5\not{\!\!K_1}$ 
($\gamma_5\not{\!\!K_2}$) we can write
\ba
& &\langle p h'|J_\mu^P|h'h\rangle = 
\frac{f^2 F_1^V}{4V^2m_\pi^2}i\epsilon_{zab}\langle t_p|\tau_a|t_{h'}\rangle
	\langle t_{h'}|\tau_b|t_h\rangle \nonumber \\ 
&\times &
\sqrt{(1+\varepsilon_1)(1+\varepsilon'_1)(1+\varepsilon_2)
(1+\varepsilon'_2)}\frac{(K_1-K_2)_\mu}{(K_1^2-m_\pi^2)(K_2^2-m_\pi^2)}
\nonumber \\
&\times & \chi^\dagger_{s'_1}\left[m_N\sigvec\cdot\kvec_1-
E_{k_1}\left(\frac{\sigvec\cdot\etavec_1}{1+\varepsilon_1}+
\frac{\sigvec\cdot\etavec'_1}{1+\varepsilon'_1}\right)+
m_N\frac{\sigvec\cdot\etavec'_1}{1+\varepsilon'_1}(\sigvec\cdot\kvec_1)
\frac{\sigvec\cdot\etavec_1}{1+\varepsilon_1}\right]\chi_{s_1}
\nonumber \\
&\times &\chi^\dagger_{s'_2}\left[m_N\sigvec\cdot\kvec_2-
E_{k_2}\left(\frac{\sigvec\cdot\etavec_2}{1+\varepsilon_2}+
\frac{\sigvec\cdot\etavec'_2}{1+\varepsilon'_2}\right)+
m_N\frac{\sigvec\cdot\etavec'_2}{1+\varepsilon'_2}(\sigvec\cdot\kvec_2)
\frac{\sigvec\cdot\etavec_2}{1+\varepsilon_2}\right]\chi_{s_2}\ .
\label{eq60}
\ea

Following the arguments discussed for the seagull current we
expand up to first order in powers of the variables
$\{\etavec_1, \etavec_2, \etavec'_2, \kvec_2\}$, whereas
$\{\etavec'_1, \kappavec, \kvec_1\}$ are treated exactly.
The non--relativistic reductions for the kinematic variables are
given by Eqs.~(\ref{eq40}--\ref{eq43}) and
after some algebra one obtains
\be
\overline{J}_{\mu}^{P}(1,2) \simeq 
-\frac{\cal F}{\sqrt{1+\lambda}}
\frac{m_N(K_1-K_2)_\mu} {(K_1^2-m_\pi^2)(K_2^2-m_\pi^2)}
\sigvec_1\cdot\kvec_1 \sigvec_2\cdot\kvec_2\ .
\label{eq61}
\ee
Using again the first--order relation 
$\lambda \simeq \tau +\kappavec\cdot\etavec_2$ and
expanding $1/\sqrt{1+\lambda}$ in powers of $\eta_2$, the pion--in--flight
matrix element may finally be cast into the form
\be
\overline{J}_{\mu}^{P}(1,2) \simeq 
-\frac{\cal F}{\sqrt{1+\tau}}
\frac{m_N(K_1-K_2)_\mu} {(K_1^2-m_\pi^2)(K_2^2-m_\pi^2)}
\sigvec_1\cdot\kvec_1 \sigvec_2\cdot\kvec_2\ .
\label{eq62}
\ee
This expression is similar to the traditional non--relativistic 
current~\cite{Alb90} except for the common 
factor $1/\sqrt{1+\tau}$, which should again include important aspects
of relativity not taken into account in the traditional non--relativistic reduction.

Once more we can express this matrix element in terms of $\kappavec$,
$\etavec_1$ and $\etavec_2$ and keep only linear terms in the small momenta,
obtaining
\ba
\overline{J}_{0}^{P}(1,2) &=& 
-\frac{4{\cal F} m_N^2}{(K_1^2-m_\pi^2)(K_2^2-m_\pi^2)}
\frac{\tau}{\sqrt{1+\tau}} 
\sigvec_1\cdot\kappavec
\sigvec_2\cdot(\etavec_1-\etavec_2)
\label{bai3}
\\
\overline{\jvec}^{P}(1,2) &=& 
-\frac{4{\cal F} m_N^2}{(K_1^2-m_\pi^2)(K_2^2-m_\pi^2)}
\frac{\sigvec_1\cdot\kappavec \sigvec_2\cdot(\etavec_1-\etavec_2)}
{\sqrt{1+\tau}} 
\kappavec\ .
\label{bai4}
\ea
Note that the space component of the pionic current is, in leading order,
purely longitudinal; its transverse components are in fact of second order in
$\eta_F$. 

Finally the limit $\eta_F\rightarrow 0$ reduces to
$\langle J_\mu^P \rangle =0$,
and  we see that all the components of the pionic current
are of $O(\eta_F)$ in the expansion.

%*********************
\section*{3. Results}
%*********************

We present here a discussion of the numerical results obtained for the MEC 
matrix elements. In the following we
compare the fully--relativistic results with those 
obtained using two different expansions for the MEC:
i) the traditional non--relativistic approach (TNR), where $\kappa$,
$\lambda$ and all nucleon three--momenta ($\eta$) are treated as being
small and ii) our new non--relativistic (also referred to as ``relativized")
approach (NR) where we expand
only in powers of bound nucleon three--momenta, not in $\kappa$ or 
$\lambda$.
In order to check the validity of our 
expansions we compute the transition matrix element of the current
between the Fermi sea and a p-h excitation, {\em i.e.}
$\langle ph^{-1}|J^{MEC}_{\mu}|F\rangle$.
Furthermore, to assess
the quality of the expansions performed for the MEC in the
last section, we show here comparisons of matrix elements for the relativistic
MEC taken between Dirac spinors with matrix elements of
the expanded MEC taken between Pauli spinors. In what follows we use
the notation of the lower diagrams shown in Fig.~1.
The connection with the general terminology introduced
in the previous section is given by the following relations:
$P=P'_1$ is the four--momentum of the ejected nucleon and
$H=P_2$, $H'=P'_2=P_1$ are the initial and
intermediate four--momenta of the bound nucleons. The corresponding
three--momenta are denoted by: $p=|{\bf p}|=|{\bf p}'_1|$,
$h=|{\bf h}|=|{\bf p}_2|$ and $h'=|{\bf h}'|=|{\bf p}'_2|=|{\bf p}_1|$.

%===================================
\subsection*{3.1. Seagull current}
%===================================

In order to simplify our analysis we first extract from the
currents the factors which are common to both relativistic and
non--relativistic  currents, namely, coupling constants, 
form factors  and isospin matrix elements. Accordingly, for the seagull current 
$J^S_{\mu}$, we define a dimensionless function $K^S_{\mu}(q,\omega,\nh)$ 
as follows:
\begin{equation}
\sum_{h'}
\langle p h'|J^{S}_{\mu}| h' h \rangle = 
\frac{f^2}{V m_\pi^2}i \langle [\ntau_1\times\ntau_2]_z \rangle 
F_1^V  k_F^2 K^S_{\mu}(q,\omega,\nh)\ ,
\label{eq67}
\end{equation}
where $\langle [\ntau_1\times\ntau_2]_z \rangle $ stands for the 
corresponding matrix element and implies a summation over isospin.
Using the definition of the seagull current, the expression for the function 
$K^S_{\mu}$ is
\begin{eqnarray}
K^S_{\mu}(q,\omega,\nh)
          &=& \frac{1}{k_F^2}
              \sum_{s_{h'}}
              \int \frac{{\rm d}^3h'}{(2\pi)^3}
              \left[ 
              \frac{\overline{u}(p)\gamma^5(\not{\!\!P}-\not{\!\!H'})u(h')}%
              {(P-H')^2-m_\pi^2}
              \overline{u}(h')\gamma_5\gamma_{\mu}u(h)
              \right. \nonumber\\
         &-&\left. 
              \frac{\overline{u}(h')\gamma^5(\not{\!\!H'}-\not{\!\!H})u(h)}%
              {(H'-H)^2-m_\pi^2}
              \overline{u}(p)\gamma_5\gamma_{\mu}u(h')
              \right],
\label{eq68}
\end{eqnarray}
where the sum runs over the third spin component of the spinor 
$u(h')\equiv u({\bf h}',s_{h'})$ 
and the integral over $\nh'$ is performed below the Fermi sea
 $|\nh'|\le k_F$.
This expression for $K^S_{\mu}$ is the exact relativistic result. 
Note that we have divided by the squared Fermi momentum $k_F^2$ in order
to obtain a dimensionless function; a factor $k_F^2$ is correspondingly
included in Eq.~(\ref{eq67}). In the previous section
we have performed an expansion in powers of the small quantities
$h/m_N$ and $h'/m_N$. Therefore we define for the components of the function
$K^S_{\mu}$ the following ``non--relativistic'' approximations :
\begin{eqnarray}
K_0^{S,NR1}(q,\omega,\nh) &=& -\frac{1}{2m_Nk_F^2}\frac{1}{\sqrt{1+\tau}}
               \sum_{s_{h'}}
              \int \frac{{\rm d}^3h'}{(2\pi)^3}
              \nonumber\\
&&            \left[ 
              \frac{%
              \chi^\dagger_{s_p}\sigvec\cdot(\np-\nh'-\tau\nh')\chi_{s_{h'}}}%
              {(P-H')^2-m_\pi^2}
              \chi^\dagger_{s_{h'}}\sigvec\cdot(\nh'+\nh)\chi_{s_h}%
              \right. 
              \nonumber\\
&&            \left. 
              -\frac{%
              \chi^\dagger_{s_{h'}}\sigvec\cdot(\nh'-\nh)\chi_{s_h}}%
              {(H'-H)^2-m_\pi^2}
	      \chi^\dagger_{s_p}\sigvec\cdot(\np+\nh'+\tau\nh')\chi_{s_{h'}}
              \right]
	\label{eq69}   \\
\nK^{S,NR1}(q,\omega,\nh) &=& -\frac{1}{k_F^2}
              \frac{1}{\sqrt{1+\tau}}
               \sum_{s_{h'}}
              \int \frac{{\rm d}^3h'}{(2\pi)^3}
              \nonumber\\
&&            \left\{
              \frac{%
              \chi^\dagger_{s_p}\nsigma\cdot\left[
              \left(1-\frac{\nkappa\cdot\nh}{2m_N(1+\tau)}\right)
              (\np-\nh')-\tau\nh'\right]\chi_{s_{h'}}}%
              {(P-H')^2-m_\pi^2}
              \chi^\dagger_{s_{h'}}\nsigma\chi_{s_h}
              \right. 
              \nonumber\\
&&            \left. -
              (1+\tau)
              \frac{%
              \chi^\dagger_{s_{h'}}\sigvec\cdot(\nh'-\nh)\chi_{s_h}}%
              {(H'-H)^2-m_\pi^2}
	      \chi^\dagger_{s_p}\nsigma\chi_{s_{h'}}
              \right\}\ ,\label{eq70}
\end{eqnarray}
where ``S,NR1'' is meant to denote non--relativistic approximation
number 1 for the seagull contributions.
These expressions should be compared with the traditional
non--relativistic seagull current that can be obtained by taking the
limit $\kappa\rightarrow 0$ and $\tau\rightarrow 0$, namely
\begin{eqnarray}
K_0^{S,TNR} (q,\omega,\nh)&=& -\frac{1}{2m_Nk_F^2}
               \sum_{s_{h'}}
              \int \frac{{\rm d}^3h'}{(2\pi)^3}
              \left[
              \frac{%
              \chi^\dagger_{s_p}\sigvec\cdot(\np-\nh')\chi_{s_{h'}}}%
              {(P-H')^2-m_\pi^2}
              \chi^\dagger_{s_{h'}}\sigvec\cdot(\nh'+\nh)\chi_{s_h}%
              \right. 
              \nonumber\\
&&            \left. \mbox{}-
              \frac{%
              \chi^\dagger_{s_{h'}}\sigvec\cdot(\nh'-\nh)\chi_{s_h}}%
              {(H'-H)^2-m_\pi^2}
	      \chi^\dagger_{s_p}\sigvec\cdot(\np+\nh')\chi_{s_{h'}}
              \right]
          \label{eq71}    \\
\nK^{S,TNR} (q,\omega,\nh)&=& -\frac{1}{k_F^2}
               \sum_{s_{h'}}
              \int \frac{{\rm d}^3h'}{(2\pi)^3}
              \left[
              \frac{%
              \chi^\dagger_{s_p}\nsigma\cdot
              (\np-\nh')\chi_{s_{h'}}
              }%
              {(P-H')^2-m_\pi^2}
              \chi^\dagger_{s_{h'}}\nsigma\chi_{s_h}
              \right. 
              \nonumber\\
&&            \left. \mbox{}-
              \frac{%
              \chi^\dagger_{s_{h'}}\sigvec\cdot(\nh'-\nh)\chi_{s_h}}%
              {(H'-H)^2-m_\pi^2}
	      \chi^\dagger_{s_p}\nsigma\chi_{s_{h'}}
              \right]\ ,\label{eq72}
\end{eqnarray}
where ``S,TNR'' denotes the traditional non--relativistic 
approximation for the seagull contributions.
In addition one would like to find approximations to the currents
where the relativistic effects are accounted for as 
corrective factors consisting of simple functions of $(q,\omega)$
such as the combination of $\kappa$, $\tau$ and $\sqrt{1+\tau}$ previously
found for the case of the single--nucleon current
(see Sect.~2.1). One could thus easily implement the relativistic 
corrections in existing models of traditional
non--relativistic MEC. For these reasons we 
define a second approximation for the seagull current in which we neglect 
the factors $\kappavec\cdot\nh/[2m_N(1+\tau)]$
and $\tau\nh'$ in the function $K^{S,NR1}$, thereby yielding a second
non--relativistic approximation for the seagull current, 
$K^{S,NR2}_{\mu}$, which differs from the traditional current only by 
the factor $\sqrt{1+\tau}$:
\begin{eqnarray}
K_0^{S,NR2} (q,\omega,\nh)&=& -\frac{1}{2m_Nk_F^2}\frac{1}{\sqrt{1+\tau}}
               \sum_{s_{h'}}
              \int \frac{{\rm d}^3h'}{(2\pi)^3}
              \left[
              \frac{%
              \chi^\dagger_{s_p}\sigvec\cdot(\np-\nh')\chi_{s_{h'}}}%
              {(P-H')^2-m_\pi^2}
              \chi^\dagger_{s_{h'}}\sigvec\cdot(\nh'+\nh)\chi_{s_h}%
              \right. 
              \nonumber\\
&&            \left. \mbox{}-
              \frac{%
              \chi^\dagger_{s_{h'}}\sigvec\cdot(\nh'-\nh)\chi_{s_h}}%
              {(H'-H)^2-m_\pi^2}
	      \chi^\dagger_{s_p}\sigvec\cdot(\np+\nh')\chi_{s_{h'}}
              \right]
              \label{eq73}\\
           &=& \frac{1}{\sqrt{1+\tau}}K^{S,TNR}_0(q,\omega,\nh)
               \label{eq74}\\
\nK^{S,NR2} (q,\omega,\nh)&=& -\frac{1}{k_F^2}
              \frac{1}{\sqrt{1+\tau}}
               \sum_{s_{h'}}
              \int \frac{{\rm d}^3h'}{(2\pi)^3}
              \left[          
              \frac{%
              \chi^\dagger_{s_p}\nsigma\cdot
              (\np-\nh')
              \chi_{s_{h'}}
              }%
              {(P-H')^2-m_\pi^2}
              \chi^\dagger_{s_{h'}}\nsigma\chi_{s_h}
              \right. 
              \nonumber\\
&&            \left. \mbox{}-
              (1+\tau)
              \frac{%
              \chi^\dagger_{s_{h'}}\sigvec\cdot(\nh'-\nh)\chi_{s_h}}%
              {(H'-H)^2-m_\pi^2}
	      \chi^\dagger_{s_p}\nsigma\chi_{s_{h'}}
              \right],\label{eq75}
\end{eqnarray}
where ``S,NR2'' denotes non--relativistic approximation
number 2 for the seagull contributions.
In what follows we check the validity of the various approximations 
to the full current introduced above by performing numerical calculations
of the functions $K^S_{\mu}(q,\omega,\nh)$ 
for several choices of the kinematical variables. 

First we notice that, for fixed momentum and energy transfer,
$(q,\omega)$, there are restrictions on the values of
the momentum of the hole $\nh$, since our nucleons are on--shell and
the momentum of the ejected particle 
$P^{\mu}=H^{\mu}+Q^{\mu}$ must satisfy $P_{\mu}P^{\mu}=m_N^2$. 
In particular, the restriction

\begin{equation}\label{restriction}
2\nh\cdot\nq = \omega^2-q^2+2E_h\omega\ ,
\end{equation}
fixing the angle between the hole momentum $\nh$ and 
the momentum transfer $\nq$ is seen to hold. Therefore the functions
$K^{S}_{\mu}$ depend only on the variables $(q,\omega,h,\phi_h)$,
where $h=|\nh|$ is the magnitude of the hole momentum and
$\phi_h$ is the azimuthal angle of $\nh$ in a coordinate system 
with the $z$-axis in the direction of $\nq$. 
The angle between $\nh$ and $\nq$ is then given by
\begin{equation}
\cos\theta_h = \frac{\omega^2-q^2+2E_h\omega}{2hq}\ .
\label{eq77}
\end{equation}
As this must lie between $-1$ and 1, one obtains a
restriction on the values of $\omega$ for which a 
contribution to the on--shell matrix element exists.

For fixed  values of $(q,\omega)$, 
there is another
restriction generated by the condition that the particle momentum 
$\np$ must lie above the Fermi sea; in this work, however, we are only
interested in the high--momentum region where relativistic corrections 
are expected to be important and there such Pauli--blocking effects can be 
ignored.

In Figs.~2--4 we show the dominant (real or imaginary) parts of the
four vector components of the seagull function $K^{S}_{\mu}$. 
The components not shown in the figures are found to be
negligible in our calculations, as a result of cancellations occurring among
different pieces: in order to understand the reasons for these cancellations 
we have explored the symmetries of the integrals involved in the various 
components. A summary of that study is given in the Appendix for the particular
case of the pion-in-flight current; a similar procedure can be
followed for the seagull current.

We choose the typical value 
$k_F=250$ MeV/c for the Fermi momentum and in all of the figures
the hole kinematics correspond to
$h=175$ MeV/c and  $\phi_h=0^{\rm o}$ . 
Because of the above mentioned symmetries in the currents it is possible to
relate the matrix elements corresponding to other choices of the angle $\phi_h$
to the ones calculated here --- 
for example the symmetry between the $\phi_h =90^{\rm o}$ and
$\phi_h =0^{\rm o}$ cases is discussed for the pion-in-flight current in 
the Appendix. We have checked that the same connection obtains for the 
values $\phi_h=90^{\rm o}$ and $\phi_h=180^{\rm o}$, with the
exception that in this case the roles of some of the components
are switched.
However, since no new information concerning the validity of the expansion is 
obtained from values of $\phi_h$ different from zero, we show 
results only for the latter case.
 
The function $K^S_{\mu}$ is displayed for three values of the momentum
transfer, namely $q=500,1000$ and 2000 MeV/c, as a function of $\omega$.
For each $q$, the allowed values of $\omega$ are restricted
to the intervals displayed in the figures. Note that the $\omega$-values 
in the figures have been chosen to lie in a region around the approximate
quasielastic peak position, $\omega=\sqrt{q^2+m_N^2}-m_N$,
which for the selected momenta occurs for
$\omega \sim 125, 433$ and 1271 MeV, respectively
(approximately at the center of the $\omega$--region displayed 
in each situation).

In each of the panels in Figs.~2--4 we plot four curves,
corresponding to the fully--relativistic 
$K^{S}_{\mu}$ (solid lines) and to the
non--relativistic approximation  $K^{S,NR1}_{\mu}$ (dashed lines); moreover the
results obtained with
the traditional non--relativistic current $K^{S,TNR}_{\mu}$ (dot-dashed
lines) and with
our simplified, non--relativistic approximation $K^{S,NR2}_{\mu}$ (dotted 
lines) are also shown. The functions $K^S_{\mu}$ are spin--matrices, 
i.e., they depend on the 
spin projections $s_p$, $s_h$ of the particle and of the hole,
respectively; accordingly we write 
\begin{equation}
K^{\mu}= \left(\begin{array}{cc} 
                K^{\mu}_{11} &
                K^{\mu}_{12} \\
                K^{\mu}_{21} &
                K^{\mu}_{22} 
               \end{array}
          \right) \ .
\label{eq78}
\end{equation}
For the sake of brevity, in the figures we show results only
for  the spin components $K_{11}$ and $K_{12}$, corresponding to $(s_p,s_h)=
(1/2,1/2)$ and $(1/2,-1/2)$ respectively.
Similar results are found for the remaining components of the currents. 
As for the dependence upon the angles $\phi_h$, 
relationships between the spin components displayed in the figures and the
remaining ones can be established and the same comments
made with respect to the dependence on $\phi_h$ are valid here as well. 

Looking at Figs.~2--4, we first note that the imaginary parts 
of $K^S_0$, $K^S_1$ and $K^S_3$ and the real part of $K^S_2$ are
not shown. Indeed, as mentioned before, they turn out to be 
negligible in comparison with the other components for all of the 
situations we have explored and therefore in the following we shall focus only 
on the remaining  four larger contributions to the current shown in 
the figures. 

Second, in all calculations we use relativistic kinematics
and the full pion propagator
even when computing the traditional matrix element.  
The importance of using relativistic kinematics is crucial
in the evaluation of the angle between $\nh$ and $\nq$
arising from the on--shell condition given in Eq.~(\ref{restriction}). 
Non--relativistic kinematics would lead instead to
the relationship $2\nh\cdot\nq=2m_N\omega-q^2$, where
a factor $\omega^2$, which is clearly important for the high values of
$q$ considered here, does not appear. In fact, as pointed in 
Refs.~\cite{Ama96,Alb90} 
in discussing the one--body responses, our approximation 
to the relativistic current is accurate only if the proper 
relativistic kinematics are used in computing the energy of
the particle with momentum $\np$. In fact the 
plane waves are then solutions
of the free Klein-Gordon equation rather than of the Schr\"odinger equation,
thus automatically accounting for relativity in the 
kinetic energy operator. The additional relativistic dynamics incorporated here,
arising from the Dirac spinology, enter as modifications of the current 
operator. For high momentum transfers, both ingredients (relativistic kinematics
and current corrections due to spinology) are of the same
level of importance. 

For moderate momentum transfers, say
$q=500$ MeV/c, the relativistic kinematics alone allow one to 
obtain agreement between the relativistic $K^S_{\mu}$
and traditional $K^{S,TNR}_{\mu}$ functions shown in Fig.~2
(compare solid with dot-dashed lines). Although the agreement becomes 
better for approximations $NR1$ and $NR2$, all of 
the curves turn out to be close enough to each other to allow the conclusion
that for moderate values of $q$ 
it is sufficient to use the traditional seagull current 
(but including relativistic kinematics) 
for computing one--particle knock--out matrix elements.

The situation changes at higher $q$-values. 
As shown in Figs.~3 and 4, the traditional approximation
(dot-dashed line) for the time and longitudinal components,
Re$K^S_0$ and Re$K^S_3$,
does not agree with the exact relativistic result (solid line).
This reflects the approximations made in the derivation of
this current, where in particular the factors $\kappa$ and $\lambda$ are
(incorrectly) treated as of higher order. In contrast, our approximations $NR1$
and  $NR2$ (dashed and dotted lines) are both very close to the
fully--relativistic result for all of the $q$-values considered. Therefore the
relativistic corrections included in these components of the current
in our approximation NR1 (or in its simplified version NR2) appear to be 
sufficient for a proper description of the relativistic effects. 

With regard to the transverse components of the seagull current in the 
$s_p =1/2$, $s_h=-1/2$ case, namely Re$K^S_1$ and Im$K^S_2$,
we first note that for low  momentum transfers they dominate over
the longitudinal components to the left of the quasielastic peak.
As functions of $\omega$ these transverse 
components are nearly linear and
cross the $\omega$-axis somewhere to the right of the
quasielastic peak. This change of sign accounts for 
the negative interference between the seagull
and one--body current contributions in the transverse electromagnetic
response to the right of the quasielastic peak~\cite{Alb90,Ama94b}.
The value of $\omega$ where these functions vanish
decreases with increasing $q$: for $q=2000$
MeV/c the position of the zero almost coincides with the center of the
quasielastic peak. The figures also show that at low energy 
(below the quasielastic peak) the exact result almost coincides 
with the other three curves. On the other hand, for high energy (above the
quasielastic peak)  
discrepancies occur between the exact and traditional currents:
thus at the end of the 
allowed $\omega$-region, the traditional current 
(dot-dashed lines) accounts for only about one-half of
the exact result (solid lines) in absolute value.
On the other hand,  our two non--relativistic
 approximations NR1 and NR2 (dashed and dotted lines) 
are much closer to the exact result. 
Hence our results show 
 that the relativized, simplified  approximation 
NR2 to the seagull current is a valid representation of
the exact relativistic current for all of the values of
the momentum transfer considered here. 

%=========================================
\subsection*{3.2. Pion--in--flight current}
%=========================================

Now we perform a similar analysis for
the pion--in--flight (or pionic) current $J^P_{\mu}$. 
First we define dimensionless functions $K^P_{\mu}$ for this current
as we did for the seagull case; thus in the matrix element
\begin{equation}
\sum_{h'}
\langle p h'|J^{P}_{\mu}| h' h \rangle = 
\frac{f^2}{Vm_\pi^2}i \langle [\ntau_1\times\ntau_2]_z \rangle 
F_1^V k_F^2 K^P_{\mu}(q,\omega,\nh)
\label{eq79}
\end{equation}
the pionic function $K^P_{\mu}$ reads
\begin{equation}
K^P_{\mu}(q,\omega,\nh)
          = \frac{1}{k_F^2}
              \sum_{s_{h'}}
              \int \frac{{\rm d}^3h'}{(2\pi)^3}
              (P+H-2H')_{\mu}
              \frac{%
              \overline{u}(p)\gamma_5(\not{\!\!P}-\not{\!\!H'})u(h')
              \overline{u}(h')\gamma_5(\not{\!\!H'}-\not{\!\!H})u(h)}
               { [(P-H')^2-m_\pi^2] [(H'-H)^2-m_\pi^2] }\ .
\label{eq80}
\end{equation}
We also introduce the traditional non--relativistic function,
denoted ``P,TNR'', for the pionic contributions
\begin{equation}
K^{P,TNR}_{\mu}(q,\omega,\nh)
          = \frac{1}{k_F^2}
              \sum_{s_{h'}}
              \int \frac{{\rm d}^3h'}{(2\pi)^3}
              (P+H-2H')_{\mu}
              \frac{%
              \chi^\dagger_{s_p}\nsigma\cdot(\np-\nh')\chi_{s_{h'}}
              \chi^\dagger_{s_{h'}}\nsigma\cdot(\nh'-\nh)\chi_{s_h}}
               {[(P-H')^2-m_\pi^2][(H'-H)^2-m_\pi^2]}\ .
\label{eq81}
\end{equation}
Finally, our approximated pionic function, introduced 
in Sect.~2, is simply given by 
\begin{equation}
K^{P,NR}_{\mu}(q,\omega,\nh) = 
\frac{1}{\sqrt{1+\tau}}K^{P,TNR}_{\mu}(q,\omega,\nh)\ ,
\label{eq82}
\end{equation}
where ``P,NR'' stands for non--relativistic approximation
for the pionic contributions.

In Figs.~5--7 we display the various components of the 
pionic function $K^P_{\mu}$ for
the same kinematics as employed for the seagull current. 
The meaning of the curves is the same as in Figs.~2--4,
except that now only one relativistic approximation is suggested, since the
simplicity of our result does not require additional assumptions.
The results obtained are similar to the
ones already found for the seagull current, and we can summarize them 
in the following points:

\begin{itemize}

\item 
The important contributions arise
from the real parts of $K^P_0$, $K^P_1$ and $K^P_3$,
and imaginary part of $K^P_2$. The reasons for this dominance
are discussed in the Appendix, where we are able to inter-relate the 
general behavior of the curves shown in Figs.~5--7 and we explore the
symmetries of the various pieces of this current.

\item 
For low momentum transfers ($q$ below some ``moderate'' value of 
about 500 MeV/c) the traditional approach, the exact matrix elements 
and, of course, our present approximation
are all very close if one takes into account relativistic kinematics. 
This justifies the use of the traditional MEC for low to moderate momentum 
transfers~\cite{Alb90,Ama94b,Ama94c}.

\item 
For high momentum transfers ($q\geq 1000$ MeV/c) the traditional expression 
(dot-dashed line) clearly disagrees
with the fully relativistic result (solid line). 
From Figs.~6 and 7 it is apparent that the
major differences between these two functions disappear if we use our
approximated current P,NR (dashed lines).
In fact, it is a particularly gratifying result 
that, apart from using relativistic kinematics, 
the simple factor $1/\sqrt{1+\tau}$
applied to the traditional current 
$K^{P,TNR}_{\mu}$ is able to reproduce the 
exact relativistic matrix element remarkably well.
Although some disagreements between our approximation
and the relativistic current for the case of Im $K^P_2$ exist, 
it is however clear that the traditional result is much worse. Moreover, 
the disagreement is found only away from the quasielastic
peak where the corresponding one--particle emission response is small,
the two matrix elements (solid and dashed lines) 
being equal at the peak 
where the approximations associated with the factor  $1/\sqrt{1+\tau}$
are expected to work better.

\end{itemize}

In conclusion we see that the new
pionic current obtained by multiplying the 
traditional non--relativistic one with the spinology factor $1/\sqrt{1+\tau}$ significantly
improves the relativistic content of the current and hence one can
use this current for computing one--particle emission responses 
for high momentum transfers within non--relativistic models, at least 
near the quasielastic peak.

%=========================================
\subsection*{3.3. The large--$q$ limit}
%=========================================

Let us end this section with a brief discussion of 
the behavior of the currents in the large--$q$ limit.
We start from the non--relativistic reductions of the
currents as given in 
Eqs.~(\ref{eq22},\ref{eq28},\ref{bai1},\ref{bai2},\ref{bai3},\ref{bai4})
and consider the limit $\kappa\to\infty$.  For these conditions 
\be
\lambda = \frac{1}{2} \left[ \sqrt{(2\kappavec+\etavec)^2+1} -
\sqrt{\etavec^2+1} \right] \simeq \kappa
\label{bai5}
\ee
and
\be
\tau = (\kappa+\lambda)(\kappa-\lambda)
\simeq \kappa (1-\hat{\kappavec}\cdot\etavec)\ .
\label{bai6}
\ee
By inserting Eqs.~(\ref{bai5},\ref{bai6}) into the currents we obtain then for
the single--nucleon current

\ba
\overline{J^0}
&\simeq&
	\sqrt{\kappa} \left[ G_E (1+\frac{1}{2}\hat{\kappavec}\cdot\etavec)
	+ i \left(G_M-\frac{G_E}{2}\right) 
        (\hat{\kappavec}\times\etavec) \cdot \sigvec \right]
\label{12}
\\
\overline{\bd J}^\parallel &=& 
\frac{\lambda}{\kappa} \overline{J^0} \hat{\kappavec}
\simeq \overline{J^0} \hat{\kappavec}
\label{13}
\\
\overline{\bd J}^\perp &\simeq&
i \sqrt{\kappa} G_M \left\{
	(\sigvec\times\hat{\kappavec}) - \frac{1}{2}  
	\left[ (\sigvec\cdot\hat{\kappavec}) (\hat{\kappavec}\times\etavec) 
       + (\sigvec\times\hat{\kappavec}) (\hat{\kappavec}\cdot\etavec) \right]
\right.
\nonumber
\\
&&
\left.
       - \frac{i}{2} (\etavec-(\hat{\kappavec}\cdot\etavec)
	\hat{\kappavec})
\right\}
\label{14}
\ea
and likewise, for the meson--exchange currents,

\ba
\overline{J}_{0}^{S}(1,2)  &\simeq& {\cal F} \frac{\sqrt{\kappa}}{m_\pi^2}
(\sigvec_1\cdot\hat{\kappavec}) \sigvec_2\cdot(\etavec_1-\etavec_2)
\label{21}
\\
\overline{\jvec}^{S}(1,2) &\simeq& {\cal F} \frac{\sqrt{\kappa}}{m_\pi^2}
\sigvec_1 \sigvec_2\cdot(\etavec_1-\etavec_2)
\label{22}
\ea

\ba
\overline{J}_{0}^{P}(1,2) &\simeq& -{\cal F} \frac{\sqrt{\kappa}}{m_\pi^2}
(\sigvec_1\cdot\hat{\kappavec}) \sigvec_2\cdot(\etavec_1-\etavec_2)
\label{23}
\\
\overline{\jvec}^{P}(1,2) &\simeq& -{\cal F} \frac{\sqrt{\kappa}}{m_\pi^2}
\hat{\kappavec}
(\sigvec_1\cdot\hat{\kappavec}) \sigvec_2\cdot(\etavec_1-\etavec_2)\ ,
\label{24}
\ea
where the inverse of the pion propagator has been expanded to leading order in
the parameters $\eta$ and $1/\kappa$:

\ba
K_1^2 - m_\pi^2
&=& m_N^2 \left[ (\varepsilon'_1-\varepsilon_1)^2 - (\etavec'_1-\etavec_1)^2 
\right] - m_\pi^2
\simeq -4 m_N^2 \kappa (1-\hat{\kappavec}\cdot\etavec_1) 
\label{19}
\\
K_2^2 - m_\pi^2
&=& m_N^2 \left[ (\varepsilon'_2-\varepsilon_2)^2 - (\etavec'_2-\etavec_2)^2 
\right] - m_\pi^2
\simeq - m_\pi^2\ .
\label{20}
\ea

At large $q$, it is of significance that:

\begin{itemize}

\item
All of the currents grow asymptotically as $\sqrt{\kappa}$. This result 
is supported by our numerical results, which show that the currents at 
$q=2000$ MeV/c are roughly twice as large as those at $q=500$ MeV/c.

\item
The time components of the seagull and pion-in-flight currents tend to cancel
each other. The same happens for the longitudinal components. Hence {\em only
the transverse components of the seagull current survives in this limit}.

\item
The longitudinal and time components of both the seagull and pion-in-flight
currents become equal as $q\to\infty$.
Since in this limit $\lambda\simeq\kappa$, it follows that
{\em in the large--$q$ limit the seagull and pionic currents are separately
gauge invariant}. Moreover, in this kinematical regime, the correlations among
nucleons are not expected to play a significant role, thus implying the
separate realization of gauge invariance in each sector of the nuclear response.
By extension, the current carried by each individual meson should be expected to
be separately conserved.

\item
Finally, if the form factors are neglected, the single--nucleon
current and the MEC display the same asymptotic behavior in $q$.
Of course the inclusion of form factors will change the $q$-dependences of the
currents. 
\end{itemize}

%************************************************************

\section*{4. Summary and Conclusions}\setcounter{footnote}{0}

%************************************************************

In this work we have found new approximations to pionic electromagnetic
meson--exchange currents using an approach which parallels recent work
involving expansions of the electroweak single--nucleon current in powers
of the  momentum of the initial bound nucleon $\eta=p/m_N$. Our goal here 
and in that previous work has been to obtain current operators that 
can be implemented in computing response functions for high momentum 
transfers in quasielastic kinematics using non--relativistic models. Our
approach allows features of relativity to be taken into account through
the use of relativistic kinematics and the  Dirac-spinology content implicit  
in the new currents. 

In this paper we have first illustrated the basic procedure by reviewing
the simpler case involving the expansion of the single--nucleon current. 
We have then turned to our main focus in the present work and applied the 
expansion ideas to a study of the pion--exchange seagull and pion--in--flight 
MEC. A distinguishing feature of the recently--obtained 
single--nucleon current is that it incorporates relativistic effects
through multiplicative factors involving the dimensionless
variables $\kappa$, $\tau$ and $\sqrt{1+\tau}$ 
(arising from the Dirac spinology) --- factors that are easy
to implement in traditional non--relativistic models. Accordingly, in our 
expansion of the MEC we have sought to 
identify corresponding factors which can embody the essential
features of the relativistic MEC. We have also examined the behavior
of the currents in the asymptotic limit where $q\to\infty$ and found that
only the seagull transverse current survives, and moreover 
that the latter and the pionic
current are separately conserved.

Finally, we have tested the quality of our approximations
by computing the transition matrix elements
$\langle ph^{-1}|J^{MEC}_{\mu}|F\rangle$ for the various
components of the currents, {\it i.e.,} for matrix elements taken 
between the ground state of a 
Fermi gas and a particle--hole excitation. We have compared the 
exact relativistic matrix elements with our 
non--relativistic approximations and with the
traditional non--relativistic expressions. 
The differences between our newly--obtained currents and the exact ones are
small even for very high momentum transfers, whereas the traditional
expressions fail at high $q$.
Due to the quality of our results, we believe  that 
these currents can very safely be used in non--relativistic models
for computing MEC effects in one--particle emission   
nuclear responses.

%************************************************************

\section*{Appendix. Symmetries and relevance
       of the various components of the currents 
                  }

%************************************************************

In this appendix we study the structure of the integrals $K^a_{\mu}$
involved in the calculations of the MEC matrix elements
to assess the relative weight 
of the different pieces into which they may be decomposed for a variety of 
kinematical conditions. In fact, in Figs.~2--7 we have only shown the 
dominant among the four components $K^a_{\mu}$ (real or imaginary part). 
The components not shown in the
figures  have been also computed and found to be small; here
we present arguments to help in understanding why they are so.

For this purpose it is sufficient to consider only the traditional
non--relativistic currents,  
since they are modified simply by multiplicative factors in our various
approximations and thus do not change our conclusions as far as the issue at
stake here is concerned.
For illustration, we restrict ourselves to the analysis of the pion-in-flight
current; the same arguments can be applied to an analysis
of the seagull current as well.

Our analysis basically amounts to studying the behavior of the
pionic function defined by Eq.~(\ref{eq81}).
To this end, we first extract the uninteresting 
constant factor $1/k_F^2(2\pi)^3$ and define the following integral
appearing in  the non--relativistic pionic current
\begin{equation}
K^{\mu}= \sum_{s_{h'}}\int d^3h' (P+H-2H')^{\mu}
\frac{\chi^\dagger_{s_p} \nsigma\cdot(\np-\nh')
\chi_{s_{h'}}\chi^\dagger_{s_{h'}} 
      \nsigma\cdot(\nh'-\nh)\chi_{s_h}}%
{[(P-H')^2-m_\pi^2][(H'-H)^2-m_\pi^2]}\ ,
\end{equation}
which depends upon the dimensionless momenta $\kappa$, $\eta$ and $\eta_F$.
We aim to identify the leading order in the expansion in $\eta$ and $\eta_F$
of the terms into which this integral splits. 
Since $\eta<\eta_F$, we shall identify $O(\eta)=O(\eta_F)$.
From this investigation we shall see how the occurrence of
cancellations rendering some matrix elements smaller than expected emerges.

We start by performing the summation over the spin of the intermediate hole
using the completeness relation 
$\sum_{s_{h'}}\chi_{s_{h'}}\chi^\dagger_{s_{h'}} = 1$ and by defining 
the spin-matrix
\begin{eqnarray} 
\Gamma & \equiv&  \sum_{s_{h'}}\nsigma\cdot(\np-\nh')
\chi_{s_{h'}}\chi^\dagger_{s_{h'}} 
      \nsigma\cdot(\nh'-\nh) \nonumber\\
&=& \nq\cdot(\nh'-\nh)-(\nh-\nh')^2 + i [\nq\times(\nh'-\nh)]\cdot\nsigma\ .
\end{eqnarray}
We shall work in the coordinate system where $\nq=q\nne_3$ and study 
the spin components $K^{\mu}_{11}$ and $K^{\mu}_{12}$ for which
\begin{eqnarray}
\Gamma_{11} & = & \nq\cdot(\nh'-\nh)-(\nh-\nh')^2
                          = q(h'-h)_3 - (\nh-\nh')^2 \\
\Gamma_{12} &=& q(h'-h)_1-iq(h'-h)_2\ .
\end{eqnarray}

\paragraph{Spin component $K^{\mu}_{11}$.}
%========================================

In this case we obtain
\begin{equation}
K^{\mu}_{11}=
\int d^3h' (P+H-2H')^{\mu}
\frac{q(h'-h)_3-(\nh-\nh')^2}{[(P-H')^2-m_\pi^2][(H'-H)^2-m_\pi^2]}\ .
\end{equation}
As a consequence, at the non--relativistic level, we have
\begin{equation}
\im K^{\mu}_{11} = 0\ .
\end{equation}
This is no longer true for the relativistic pionic contribution, although 
the corresponding imaginary parts of the relativistic pionic 
matrix elements remain very small.

Next we examine the real components for $\mu=0,\ldots,3$.

\vspace{0.15cm}
\begin{center}
\underline{\sl  $\re K^{3}_{11}$ component}
%==========================================
\end{center}
\vspace{0.15cm}

First, the longitudinal component turns out to read
\begin{eqnarray}
\re K^{3}_{11} &=&
\int d^3h' (q+2h_3-2h'_3)
\frac{q(h'-h)_3-(\nh-\nh')^2}{[(P-H')^2-m_\pi^2][(H'-H)^2-m_\pi^2]}
\nonumber\\
&=& O(\eta_F)+O(\eta_F^2)+O(\eta_F^3)\ ,
\label{K311}
\end{eqnarray}
where $O(\eta_F)$ means that the associated integrand is
linear in $h$ or $h'$. Similarly $O(\eta_F^2)$ means that the integrand
is proportional to $h^2$, $(h')^2$ or $hh'$ and so on.
The leading terms in Eq.(\ref{K311}) are clearly of order $\eta_F$ and
$\eta_F^2$ and their expressions are given by
\begin{eqnarray}
O(\eta_F)   &=& \int d^3h' 
              \frac{q^2(h'_3-h_3)}{[(P-H')^2-m_\pi^2][(H'-H)^2-m_\pi^2]}\\
O(\eta_F^2) &=& -\int d^3h' 
              \frac{2q(h_3-h'_3)^2+q(\nh-\nh')^2}
               {[(P-H')^2-m_\pi^2][(H'-H)^2-m_\pi^2]}\ .
\end{eqnarray}
In comparing these two pieces, we see that the integrand in the 
term $O(\eta_F)$ is proportional to $h'_3-h_3$, which can be
positive or negative, thus potentially leading to cancellations. 
In fact, this term is found to be close to zero near the
quasielastic peak (QEP). The reason is the following: at the QEP 
$p=q$, and, for $q$ large, $\nh$ is almost perpendicular to $\nq$ and thus
for the case $\phi_h=0^{\rm o}$ at the QEP we can set $\nh \sim h \nne_1$.
The corresponding integral accordingly vanishes to first order, namely
\begin{equation}
\int d^3h'\frac{(h'_3-h_3)}{[(P-H')^2-m_\pi^2][(H'-H)^2-m_\pi^2]}
\simeq  \int d^3h'\frac{h'_3}{[Q^2-m_\pi^2][(H'-H)^2-m_\pi^2]} = 0 \ ,
\end{equation}
because the denominator is invariant with respect to the
inversion of $h'_3$.
As  a consequence, for the longitudinal component of the pionic current 
$K_{11}$, we have
\begin{equation} 
\re K^3_{11} \simeq 0   \kern 1cm \mbox{\rm near the QEP,}
\end{equation}
as we have checked numerically. For instance, 
in Fig.~6  we see that $\re K^3_{11}$
crosses the $\omega$-axis somewhat short of the middle of the 
$\omega$-allowed domain (the approximate position of
the QEP) because of the approximation made in the
denominator and of the presence of the other term $O(\eta_F^2)$.
This is found to be negative in this region and not entirely negligible;
hence the total matrix element reaches  zero slightly to the left of the  QEP.

Therefore, we see in this case that,
although {\em a priori} this current if of 
$O(\eta_F)$, its actual weight depends upon the value of the 
coefficient that multiplies $\eta_F$ in the expansion, which is 
$\omega$-dependent and in some cases may be small. 

On the other hand, we see in the same figure that  in the regions far 
from the QEP (especially for high $\omega$)
this component is large: indeed here the cancellations are much weaker
and then the behavior of $\re K^3_{11}$ is again of $O(\eta_F)$.

\vspace{0.15cm}
\begin{center}
\underline{\sl  $\re K^{0}_{11}$ component}
\end{center}
\vspace{0.15cm}
%=======================================

The same conclusions obtained for the third component are valid as well for the
$\re K^0_{11}$. In this case:
\begin{eqnarray}
\re K^{0}_{11} &=&
\int d^3h' \left(\omega+ 2\frac{h^2}{2m_N}-2\frac{h'{}^2}{2m_N}\right)
\frac{q(h'_3-h_3)-(\nh-\nh')^2}{[(P-H')^2-m_\pi^2][(H'-H)^2-m_\pi^2]} 
\nonumber\\
&=&  O(\eta_F)+O(\eta_F^2)+O(\eta_F^3)\ ,
\end{eqnarray}
with
\begin{eqnarray}
O(\eta_F)   &=& \int d^3h' 
              \frac{\omega q(h'_3-h_3)}{[(P-H')^2-m_\pi^2][(H'-H)^2-m_\pi^2]}\\
O(\eta_F^2) &=& -\int d^3h' 
              \frac{\omega(\nh-\nh')^2}
               {[(P-H')^2-m_\pi^2][(H'-H)^2-m_\pi^2]}\ .
\end{eqnarray}
Now, for the same reasons as before, the term $O(\eta_F)$ almost vanishes
near the QEP. Then
\begin{equation} 
\re K^0_{11} \simeq 0   \kern 1cm \mbox{\rm near the QEP.}
\end{equation}

Hence $\re K^0_{11}$ behaves like $\re K^3_{11}$, as also found
in our calculations (see Fig.~6).
Even finer details of the results can be interpreted in the same 
manner. For instance:
\begin{itemize} 

\item 
The zero of $K_{11}^0$ occurs slightly to the right of the zero of $K_{11}^3$,
because, in the present case, a piece proportional to $2q(h_3-h'_3)$ 
is missing in the term $O(\eta_F^2)$. Hence the $O(\eta_F^2)$ term is less
negative than in the $K_{11}^3$ case. 

\item Also, to second order, the following relation
\begin{equation}
K_{11}^3 = \frac{q}{\omega}K_{11}^0  -\int d^3h' 
              \frac{2q(h_3-h'_3)^2}{[(P-H')^2-m_\pi^2][(H'-H)^2-m_\pi^2]}
\label{109}
\end{equation} 
is seen to hold.
For instance, for $q=1000$ MeV/c, $\phi_h=0^{\rm o}$ and $\omega=550$ MeV, 
it turns out that
$K_{11}^0= -15$, $\frac{q}{\omega}K_{11}^0= -27$, while 
$K_{11}^3=-40$. Hence an estimate of $-13$ follows for the magnitude
of the second--order term, represented by the integral on the 
right--hand side of Eq.~(\ref{109}).
\end{itemize}

\vspace{0.15cm}
\begin{center}
\underline{\sl  $\re K^{1}_{11}$, $\re K^{2}_{11}$ components}
\end{center}
\vspace{0.15cm}
%=======================================

Concerning the transverse part we have
\begin{equation}
\re\nK^{T}_{11}=
\int d^3h' 2(\nh-\nh')_T
\frac{q(h'_3-h_3)-(\nh-\nh')^2}{[(P-H')^2-m_\pi^2][(H'-H)^2-m_\pi^2]}
= O(\eta_F^2)+O(\eta_F^3)\ .
\end{equation}
The major contribution is expected to arise from the second--order term, namely
\begin{equation}
O(\eta_F^2)=
\int d^3h' 2(\nh-\nh')_T
\frac{q(h'_3-h_3)}{[(P-H')^2-m_\pi^2][(H'-H)^2-m_\pi^2]}\ .
\end{equation}
Now, for $\phi_h=0^{\rm o}$ and near the QEP, we have $\nh\simeq h_1\nne_1$
and for the components 1 and 2 we obtain
\begin{eqnarray}
\re K^1_{11}&\simeq& 
\int d^3h' \frac{2q(h_1-h'_1)h'_3}{[Q^2-m_\pi^2][(H'-H)^2-m_\pi^2]}\simeq 0\\
\re K^2_{11}&\simeq& 
-\int d^3h' \frac{2qh'_2h'_3}{[Q^2-m_\pi^2][(H'-H)^2-m_\pi^2]}\simeq 0\ .
\end{eqnarray}
In the first case exact cancellations occur when 
$h'_3 \rightarrow -h'_3$, but in the second case we actually have 
{\em double} 
cancellations when $h'_3\rightarrow -h'_3$ and $h'_2\rightarrow-h'_2$.
Accordingly it should be expected that: 
\begin{itemize}
\item One has
\begin{equation} |\re K^2_{11}| \ll |\re K^1_{11}| \ .
\end{equation}
Actually the cancellations in $\re K^2_{11}$ are so strong that this component 
is even smaller than $\im K^2_{11}$ for the relativistic 
current (hence $\re K^2_{11}$ is not displayed). 
\item Both $\re K^2_{11}$ and $\re K^1_{11}$ vanish around the QEP. 
\item If $\phi_h=90^{\rm o}$ then $\nh \simeq h_2 \nne_2$: accordingly the roles
of $\re K^1$ and $\re K^2$ switch, i.e. 
\begin{equation} |\re K^1_{11}| \ll |\re K^2_{11}| \ .
\end{equation}
\end{itemize}
All of these properties have been checked in our numerical results.

Since $\re K^1_{11}$ is zero around the QEP for $\phi_h=0^{\rm o}$,
we infer that this piece 
is actually of $O(\eta_F^3)$. Moreover, as the approach to zero of 
$\re K^2_{11}$ is faster than in the case of $\re K^1_{11}$, $\re K^2_{11}$ is
likely of $O(\eta_F^4)$ (very small). Clearly a precise determination of the
actual order would require a more 
detailed analysis of the integrals, or to compute analytically the integrals 
in the static limit. However, we believe that the arguments given above are
enough for reaching an adequate understanding of the results.

%=========================================
\paragraph{Spin component $K^{\mu}_{12}$.}
%=========================================

For the $s_p=1/2$, $s_h=-1/2$ component we have
\begin{equation}
K^{\mu}_{12}=
\int d^3h' (P+H-2H')^{\mu}
\frac{q(h'-h)_1-iq(h'-h)_2}{[(P-H')^2-m_\pi^2][(H'-H)^2-m_\pi^2]}\ .
\end{equation}
Hence  the real and imaginary parts of this matrix element are given by
\begin{eqnarray}
\re K^{\mu}_{12}
&=& \int d^3h' (P+H-2H')^{\mu}
    \frac{q(h'-h)_1}{[(P-H')^2-m_\pi^2][(H'-H)^2-m_\pi^2]} \\
\im K^{\mu}_{12}
&=& \int d^3h' (P+H-2H')^{\mu}
     \frac{q(h-h')_2}{[(P-H')^2-m_\pi^2][(H'-H)^2-m_\pi^2]}\ .
\end{eqnarray}

\vspace{0.15cm}
\begin{center}
\underline{\sl  $K^{0}_{12}$ component}
\end{center}
\vspace{0.15cm}
%=======================================

For the time component we have
\begin{eqnarray}
K^{0}_{12} &=&
\int d^3h'
\left(\omega+ 2\frac{h^2}{2m_N}-2\frac{h'{}^2}{2m_N}\right)
\frac{q(h'-h)_1-iq(h'-h)_2}{[(P-H')^2-m_\pi^2][(H'-H)^2-m_\pi^2]}
\nonumber \\
&=& O(\eta_F)+O(\eta_F^3)\ .
\end{eqnarray}
To first order we obtain
\begin{equation}
K^{0}_{12}\simeq
\int d^3h'
\omega
\frac{q(h'-h)_1-iq(h'-h)_2}{[(P-H')^2-m_\pi^2][(H'-H)^2-m_\pi^2]}\ .
\end{equation}
For $\phi_h=0^{\rm o}$ and near the QEP again
$\nh \simeq h_1\nne_1$, and hence
\begin{equation}
K^{0}_{12}\simeq
\int d^3h'
\omega\frac{q(h'-h)_1-iqh'_2}{[Q^2-m_\pi^2][(H'-H)^2-m_\pi^2]}\ .
\end{equation}
As in the case of the $K_{11}$ component, cancellations occur 
in a way that yields 
\begin{equation}
\im K^{0}_{12}\simeq 0  \kern 1cm \mbox{\rm at the QEP.}
\end{equation}
Hence $\re K^0_{12}$ is expected to be the dominant component.

\vspace{0.15cm}
\begin{center}
\underline{\sl  $K^{3}_{12}$ component}
\end{center}
\vspace{0.15cm}
%=======================================

A similar result obtains for the longitudinal component, namely
\begin{equation}
K^{3}_{12}\simeq
\int d^3h'
q \frac{q(h'-h)_1-iq(h'-h)_2}{[(P-H')^2-m_\pi^2][(H'-H)^2-m_\pi^2]}
\end{equation}
and, for the same reasons as before,
\begin{equation}
\im K^{3}_{12}\simeq 0  \kern 1cm \mbox{\rm at the QEP.}
\end{equation}
Also the following relationship 
\begin{equation}
K^3_{12} \simeq \frac{q}{\omega}K^0_{12}
\end{equation}
is found to hold to leading order,
as can be verified in the figures. For instance, from Fig.~6, for 
$q=1000$ MeV/c and $\phi_h=0^{\rm o}$, at the 
maximum ($\omega\sim 450$ MeV) we get $\re K^{0}_{12}\sim -11$, 
$\re K^3_{12}\sim -25$ and $\frac{q}{\omega}K^0_{12}= -24.4$.

\vspace{0.15cm}
\begin{center}
\underline{\sl $K^{1}_{12}$, $K^{2}_{12}$ components}
\end{center}
\vspace{0.15cm}
%=======================================

For the transverse components we have
%--------------
\begin{eqnarray}
K^{1}_{12}&=&
\int d^3h' 2(h-h')_1
 \frac{q(h'-h)_1-iq(h'-h)_2}{[(P-H')^2-m_\pi^2][(H'-H)^2-m_\pi^2]}\\
K^{2}_{12}&=&
\int d^3h' 2(h-h')_2
 \frac{q(h'-h)_1-iq(h'-h)_2}{[(P-H')^2-m_\pi^2][(H'-H)^2-m_\pi^2]}\ ,
\end{eqnarray}
%------------
and for the real and imaginary parts of $K^1$
%---------------
\begin{eqnarray}
\re K^{1}_{12}&=&
-\int d^3h' 
 \frac{2q(h_1'-h_1)^2}{[(P-H')^2-m_\pi^2][(H'-H)^2-m_\pi^2]}\\
\im K^{1}_{12}&=&
\int d^3h' 
 \frac{2q(h_1'-h_1)(h'_2-h_2)}{[(P-H')^2-m_\pi^2][(H'-H)^2-m_\pi^2]}\ .
\end{eqnarray}
%------------
First we see that the integrand in $\re K^1_{12}$ is proportional to 
$2(h'_1-h_1)^2 >0$, and hence no cancellations occur. 
Accordingly this integral, although  of $O(\eta_F^2)$,
turns out to be of the same order of magnitude as 
$K^0_{12}$ (of $O(\eta_F)$, but with cancellations), as
is clearly seen in our results. 
On the other hand, cancellations do occur in $\im K^1_{12}$; hence, as before,
%---------------
\begin{equation}
\im K^1_{12} \simeq 0 \kern 1cm \mbox{\rm at the QEP.}
\end{equation}
%---------------
We thus expect $\re K^1_{12}$ to be the dominant part.

In connection with $K^2_{12}$ we obtain
%---------------
\begin{eqnarray}
\re K^{2}_{12}&=&
-\int d^3h' 
 \frac{2q(h_1'-h_1)(h'_2-h_2)}{[(P-H')^2-m_\pi^2][(H'-H)^2-m_\pi^2]}\\
\im K^{2}_{12}&=&
\int d^3h' 
 \frac{2q(h_2'-h_2)^2}{[(P-H')^2-m_\pi^2][(H'-H)^2-m_\pi^2]}\ .
\end{eqnarray}
%-------------
First we see that, as found in our calculations, the following relation 
%---------------
\begin{equation}
\re K^2_{12} = - \im K^1_{12}
\end{equation}
%-------------
holds, and hence the real part of $K^2_{12}$ is
negligible with respect to the imaginary part, which is thus the dominant one.
Also, since the integrand for $\im K^2_{12}$ is proportional to 
$(h_2-h'_2)^2>0$, no direct relationship between 
$\im K^2_{12}$ and $\re K^1_{12}$ exists. 

Again, for $\phi_h=90^{\rm o}$, the roles of $K^1$ and $K^2$
are switched, and from the above expressions we find
%---------------
\begin{eqnarray}
\re K^1_{12}(\phi_h=0^{\rm o}) &=& -\im K^2_{12}(\phi_h=90^{\rm o})\\
\im K^2_{12}(\phi_h=0^{\rm o}) &=& -\re K^1_{12}(\phi_h=90^{\rm o})\ .
\end{eqnarray}
%-------------

\vspace{0.5in}

{\Large \bf Acknowledgements}

\vspace{0.3in}

This work was supported in part by funds provided by 
DGICYT (Spain) under contract Nos. PB95--1204, PB95--0123 and PB95--0533--A
and the Junta de Andaluc\'\i a (Spain) and in part
by the U.S. Department
of Energy (D.O.E.) under cooperative research agreement 
\#DF-FC02-94ER40818, by the INFN-MIT ``Bruno Rossi'' Exchange Program,
and by a NATO Collaborative Research Grant Number 940183.

\vspace{1.0in}

%Could add the following references:

%\vspace{0.2in}

%J.W. Van Orden, T.W. Donnelly, T. deForest, Jr. and W.C. Hermans,
%Phys. Lett. {\bf 76B} (1978) 393.

%\vspace{0.2in}

%J.W. Van Orden and T.W. Donnelly, Ann. Phys. {\bf 131} (1981) 451.

%\vspace{0.2in}

%+ MEC references from Torino work (Maria can supply these).

%\vspace{1.0in}

%========================================================

\end{document}